\documentclass[%
 reprint,
superscriptaddress,
 amsmath,amssymb,
 aps,
 pra,
 longbibliography,
]{revtex4-2}

\usepackage[english]{babel}

\usepackage{graphicx}% Include figure files
\usepackage{dcolumn}% Align table columns on decimal point
\usepackage{bm}% bold math
\usepackage{subcaption} 
\usepackage{xcolor}%удалить потом
\usepackage{comment}%удалить потом

\makeatletter

\newcommand{\ket}[1]{\ensuremath{|{#1}\rangle}}
\newcommand{\bra}[1]{\ensuremath{\langle{#1}|}}

\newcommand{\cnj}[1]{{#1}^{\ast}}

\newcommand{\Tr}{\mathop{\rm Tr}\nolimits}

 \newcommand{\ind}[1]{\mathrm{#1}}

\newcommand{\dd}{\mathrm{d}}

\makeatother

\begin{document}
\DeclareGraphicsExtensions{.jpg,.pdf}

\preprint{APS/123-QED}

\title{Controlling Hong-Ou-Mandel antibunching via parity governed local
  spectral shaping of biphoton states}% Force line breaks with \\%

\author{Mikhail Guselnikov}\email{Corresponding author: msguselnikov@itmo.ru}
\affiliation{%
  Laboratory of Quantum Processes and Measurements, ITMO University,\\
  3b Kadetskaya Line, Saint Petersburg 199034, Russia }%
  \affiliation{%
  SMARTS-Quanttelecom LLC, \\
6th Vasilyevskogo Ostrova Line, 59, Saint Petersburg 199178, Russia}%
\author{Alexei D. Kiselev}%
\affiliation{%
  Laboratory of Quantum Processes and Measurements, ITMO University,\\
  3b Kadetskaya Line, Saint Petersburg 199034, Russia }%
\affiliation{%
  Laboratory for Quantum Communications, ITMO University, \\
  Birzhevaya Line, 16, Saint Petersburg 199034, Russia }%
  
\author{Andrei Gaidash} \affiliation{%
  Laboratory of Quantum Processes and Measurements, ITMO University,\\
  3b Kadetskaya Line, Saint Petersburg 199034, Russia }%
\affiliation{%
  SMARTS-Quanttelecom LLC, \\
6th Vasilyevskogo Ostrova Line, 59, Saint Petersburg 199178, Russia}%

\author{George Miroshnichenko} \affiliation{%
  Waveguide Photonics Research Center, ITMO University, 49 Kronverksky
  Pr., Saint Petersburg 197101, Russia }%
\affiliation{%
  Institute “High School of Engineering,” ITMO University, 49
  Kronverksky Pr., Saint Petersburg 197101, Russia }%

\author{Anton Kozubov} \affiliation{%
  Laboratory of Quantum Processes and Measurements, ITMO University,\\
  3b Kadetskaya Line, Saint Petersburg 199034, Russia }%
\affiliation{%
  SMARTS-Quanttelecom LLC, \\
6th Vasilyevskogo Ostrova Line, 59, Saint Petersburg 199178, Russia}%

\date{\today}% It is always \today, today,

\begin{abstract}
    We investigate into experimentally detectable effects determined by
the fundamental interplay between the local spectral and entanglement
properties of biphoton states such as the two-photon interference at
beam splitter phenomena known as the Hong-Ou-Mandel (HOM) bunching and
antibunching.
These regimes can be characterized using
the symmetry degree parameter $D_S$ that enters
the two-photon coincidence probability
$P_{2c}=(1-D_S)/2$.
In the case of HOM bunching (antibunching),
$D_S$ is positive (negative).
Though the symmetry degree can generally be expressed in terms of
the difference between the contributions coming from the symmetric and antisymmetric parts of
the biphoton joint spectral amplitude (JSA), $\psi(\omega_1,\omega_2)$,
for a certain physically realizable class of the JSA,
where $\psi(\omega_1,\omega_2)$ is proportional to the product of %\textcolor{magenta}{local}
amplitudes $\varphi_1(\omega_1)\varphi_2(\omega_2)$
multiplied by a Gaussian-shaped entangling factor dependent on
the frequency difference $\omega_1+\omega_2-2\Omega$ 
($2\Omega$ is the central frequency), 
we find the sign of $D_S$ is primarily governed
by the parity properties of the %\textcolor{magenta}{local}
spectral function,
$\varphi_{12}(\omega)=\varphi_1(\omega)\varphi_2^*(\omega)$.
It is the even (odd) part of $\varphi_{12}=\varphi_{12}^{(+)}+\varphi_{12}^{(-)}$
that meets the parity condition
$\varphi_{12}^{(+)}(\omega-\Omega)=\varphi_{12}^{(+)}(\Omega-\omega)$
($\varphi_{12}^{(-)}(\omega-\Omega)=- \varphi_{12}^{(-)}(\Omega-\omega)$)
to yield the positive (negative) contribution,
$D_S^{(+)}$ ($-D_S^{(-)}$),
to the symmetry degree parameter: $D_S=D_S^{(+)}-D_S^{(-)}$.
In particular,  the regime of antibunching with  $D_S=-D_{S}^{(-)}$,
which is known to take place only for entangled states,
occurs at $\varphi_{12}=\varphi_{12}^{(-)}$.
We have shown that
switching between the bunching
and antibunching regimes
can be realized using
the experimentally accessible family of modulated biphoton states
produced using the  spectral phase modulation
fine-tuned via either the sub-nanometer scale variation of the path length 
or, equivalently, the time-delay parameter of the order of attoseconds.
For this class of modulated states,
the Schmidt decomposition technique is employed  to
compute the Schmidt number as a function of the modulation parameter.
This dependence reveals the structure of narrow resonance peaks
strongly correlated with the corresponding narrow dips of
the symmetry degree
(the peaks of the two-photon coincidence probability)
where the perfect HOM antibunching occurs.

\end{abstract}

\maketitle

%\tableofcontents

%%%%%%%%%%%%%%%%%%%
\section{Introduction}
\label{sec:level1}
%%%%%%%%%%%%%%%%%%%%

Quantum technologies are rapidly transitioning from theoretical
concepts to real-world applications. While quantum computing is set to
revolutionize information processing, quantum sensing and metrology
promise unprecedented precision in measurements. In recent years,
significant research efforts have been devoted to the development of
quantum sensors, exploiting a variety of quantum processes, including
the spin dynamics of alkali atoms~\cite{Bonizzoni2024, Budker2007},
ion trapping~\cite{Biercuk2010, Campbell2017}, superconducting
dynamics in magnetic fields~\cite{Hatridge2011, Degen2017},
the Sagnac effect~\cite{Anandan1981, Bertocchi2006, Marletto2021}, and the Hong-Ou-Mandel (HOM)
effect~\cite{Hong1987} among other physical phenomena.  

HOM interference is now widely regarded as a cornerstone technology in quantum optical metrology and
sensing~\cite{Alodjants2024}. Over the past decade, numerous studies have explored the fundamental
limits and practical applicability of the HOM effect across a broad range of tasks, including
ultrasensitive measurements, quantum state tomography, spectroscopy, imaging , and
  even the development of quantum gravity theories~\cite{Douce2013, Parniak2018,Devaux2020,
    Scott2021, Fabre2021,Jordan2022,Barzel2022, Descamps2023,Meskine2024, Descamps2025}.
  One of the most spectacular achievements is the
  demonstration of the HOM interference
  with attosecond-level sensitivity to time delays approaching the quantum Cram\"er-Rao
  bound~\cite{Lyons2018, Chen2019, Fink2019,Chen2023}.

Applications of the HOM interference in quantum metrology are often based on the
  experimental observation of the resonance-like dip in the two-photon coincidence statistics called the
  HOM dip, whose minimum, where the coincidences are completely suppressed,
corresponds to the limiting case of the perfect HOM bunching.
The opposite regime of the two-photon interference at a beamsplitter~---~the so-called HOM antibunching~---~occurs when
the coincidence probability exceeds 50\%.
This regime has also been shown to offer promising sensing capabilities. It was
  theoretically predicted~\cite{Wang2006} that the HOM antibunching takes place only for entangled
  bosonic states, suggesting its potential use as a means for assessing entanglement of photon
pairs~\cite{Silberhorn2008,Fedrizzi2009,Silberhorn2013,Barbieri:scirep:2017}.
For quantum optical gyroscopes
  (QOGs) based on various combinations of the Sagnac and the HOM interferometers,
  this idea was subsequently explored experimentally in Refs.~\cite{Restuccia2019,Toros2022,Cromb2023} to
  study how 
  quantum entanglement affects the HOM interference of biphoton states in non-inertial frames. In
  the experiment
  reported in Ref.~\cite{Cromb2023}, it was demonstrated that spatial rotations of a QOG can
  induce transitions in the two-photon coincidence statistics between the HOM bunching and
  antibunching. Since the HOM antibunching occurs only if the states are entangled, it was concluded that
  spatial rotations could influence the entanglement properties of quantum states.

In this paper, we adapt a systematic approach and theoretically examine both the HOM antibunching
effect and the entanglement degree of
certain spectrally entangled
biphoton states generated via
spontaneous parametric down-conversion
(SPDC).
Our purpose is twofold: (1) to understand the key spectral properties underlying the
correlations between the HOM antibunching and the degree of entanglement, and (2) to reveal the
precise manner in which these properties and correlations can be utilized for either entanglement
control or interferometric measurements.

The paper is structured as follows.
In Sec.~\ref{subsec:biphoton-states-at-bs},
we formulate the problem of the correlation between antibunching and
quantum entanglement in biphoton states
described in terms 
of the joint
spectral amplitude (JSA)
and characterize the regimes of the HOM interference
using the symmetry degree parameter. We introduce the family of
  specifically shaped SPDC-biphotons and, for this family, derive the parity conditions giving
the spectral properties of JSA that ensure the observation of
the HOM (anti)bunching in Sec.~\ref{subsec:parity-cond}. 
 Then,
in Sec.~\ref{subsec:parity-breaking},
we discuss particular examples of experimentally realizable biphoton
states that can possess the required properties.
Section~\ref{sec:sec3} analyzes the entanglement
properties of the standard Gaussian biphotons produced via spontaneous parametric
down-conversion (SPDC) using the Schmidt decomposition
technique. Though these states cannot demonstrate antibunching,
their analytical tractability renders them useful for benchmarking
more complex states capable of exhibiting antibunching.
In Sec.~\ref{sec:sec4},
we discuss the generation and properties of biphotons of a certain kind
  referred to as the modulated biphoton states.
For these states, we provide a quantitative assessment of the sensitivity of the antibunching
effect to the degree of entanglement and discuss its potential
applications in high-precision measurements and quantum entanglement
control.

Finally, in Sec.~\ref{sec:disc-concl},
we draw our results together and make some concluding remarks.
Technical details of some caculations  are relegated to Appendices~\ref{apendA0}~--~\ref{apendB}.

%%%%%%%%%%%%%%%%%%%%%%%%%%%%%%%%%%%%%%%
\section{Hong-Ou-Mandel antibunching}
\label{sec:sec2}
%%%%%%%%%%%%%%%%%%%%%%%%%%%%%%%%%

%%%%%%%%%%%%%%%%%%%%%%%%%%%%%%%
\subsection{Interference of biphoton state at beamsplitter}
\label{subsec:biphoton-states-at-bs}
%%%%%%%%%%%%%%%%%%%%%%%%%%%%%%%%%%%%%%

\begin{figure}
\includegraphics[width=1\linewidth]{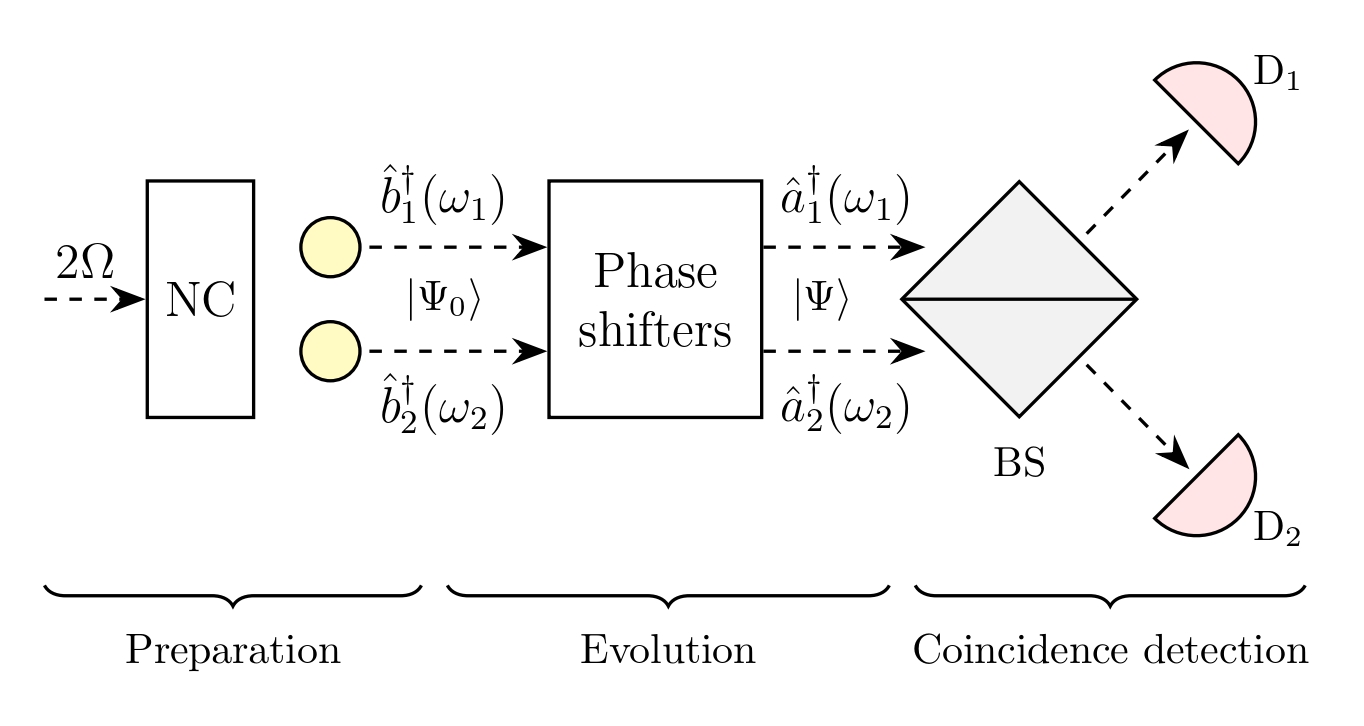}
% Here is how to import EPS art
\caption{\label{fig:fig1} Conceptual scheme of the HOM interferometer under consideration. A nonlinear crystal (NC) is pumped by the radiation with
  central frequency $2\Omega$ in order to produce a biphoton state. At
  the output of the NC, the photons are separated spatially and
  directed to the different optical-fiber channels, characterized by
  creating operators $\hat{b}^{\dagger}_1(\omega_1)$ and
  $\hat{b}^{\dagger}_2(\omega_2)$, respectively. The biphoton state
  passes through an arbitrary scheme consisting of phase shifters. After the scheme, the biphoton state
  interferes at the output beamsplitter BS and is detected by
  photodetectors $D_1$ and
$D_2$.}
\end{figure}

We consider a general optical configuration of a HOM interferometer
depicted in Fig.~\ref{fig:fig1}. A nonlinear crystal (NC) is pumped by
radiation with a central frequency of $2\Omega$ generating correlated
photon pairs via the SPDC process.  At the output of the NC, the
generated photons are spatially separated and directed into two distinct
optical fiber channels, which are characterized by the bosonic creation
operators $\hat{b}^{\dagger}_1(\omega_1)$ and
$\hat{b}^{\dagger}_2(\omega_2)$, respectively.
At the input of the fiber channels,
the photons are in
the frequency-entangled biphoton state
given by~\cite{Shih2020}
\begin{align}
  \left|\Psi_0\right\rangle = \int_{-\infty}^{\infty}d\omega_1d\omega_2
  \psi_0\left(\omega_1,
  \omega_2\right)\hat{b}^{\dagger}_{1}\left(\omega_1\right)\hat{b}^{\dagger}_{2}\left(\omega_2\right)\left|0\right\rangle,   
\label{eq:eq2.1.1}
\end{align}
where $\psi_0(\omega_1, \omega_2)$ is the initial biphoton
spectrum, which is also referred to as the  normalized joint spectral amplitude (JSA) and
meets the normalization condition:
$\int|\psi_0(\omega_1, \omega_2)|^2\dd\omega_1\dd\omega_2=1$.

The biphoton state then propagates through a generic
setup composed of phase shifters. After traversing the
phase shifters, the biphoton state undergoes interference at
the output BS and is subsequently detected by photodetectors $D_1$ and
$D_2$.

We consider that the biphoton state at the input of the final BS
can be taken in the following form:
\begin{align}
  \ket{\Psi}
  %\left|\Psi\right\rangle
  = \int_{-\infty}^{\infty}d\omega_1d\omega_2 \psi\left(\omega_1,
  \omega_2\right)\hat{a}^{\dagger}_{1}\left(\omega_1\right)\hat{a}^{\dagger}_{2}\left(\omega_2\right)\left|0\right\rangle, 
\label{eq:eq2.1.3}
\end{align}
where $\psi\left(\omega_1, \omega_2\right)$ is the JSA modified by the
phase shifters; operators
$a^{\dagger}_{1,2}(\omega_{1,2})$ ($a_{1,2}(\omega_{1,2})$)
are the creation (annihilation) operators describing the two
spatial modes of the photons entering the corresponding
two input arms
(the photons are assumed to be in the same polarization mode generated
via a type-I SPDC process) and
$\left|0\right\rangle=\left|0\right\rangle_1\otimes\left|0\right\rangle_2$,
where $\ket{0}_{1}$ and $\ket{0}_{2}$ are the vacuum states of the first
and second spatial modes, respectively.
The bosonic operators satisfy the commutation relations:
$[a_i(\omega_m),a_j^{\dagger}(\omega_n)]=\delta(\omega_m-\omega_n)\delta_{ij}$,
where $\delta_{ij}$ is the Kronecker symbol and
$\delta(\omega_m-\omega_n)$ is the Dirac delta function,
and the normalization condition for
 the biphoton state is given by
\begin{align}
\langle \Psi|\Psi\rangle = \int_{-\infty}^{\infty}d\omega_1d\omega_2 \left|\psi\left(\omega_1, \omega_2\right)\right|^2=1.
\label{eq:eq2.1.4}
\end{align}
The probability of a two-photon coincidence event, $P_{2c}$, can be represented by the well-known expression
\begin{align}
P_{2c}=\frac{1}{2}-\frac{1}{2}\int_{-\infty}^{\infty}d\omega_1d\omega_2\psi(\omega_1,
  \omega_2)\psi^*(\omega_2, \omega_1). 
\label{eq:eq2.1.11}
\end{align}
Derivation of this expression is detailed in Appendix \ref{apendA0}.

Since the JSA
$\psi(\omega_1,\omega_2)$
can be written as a sum of
the functions
$\psi_{S}\left(\omega_1, \omega_2\right)=\left[\psi\left(\omega_1,
    \omega_2\right)+\psi\left(\omega_2, \omega_1\right)\right]/2$
and
$\psi_{AS}\left(\omega_1, \omega_2\right)=\left[\psi\left(\omega_1,
    \omega_2\right)-\psi\left(\omega_2, \omega_1\right)\right]/2$
representing
the JSA components that are
symmetric and antisymmetric with respect to the permutation of arguments
$\omega_1\leftrightarrow\omega_2$, respectively,
Eq.~\eqref{eq:eq2.1.11} can be conveniently rewritten in the
alternative form:
\begin{align}
  &
    P_{2c}=\frac{1}{2}-\frac{1}{2}\times
    \notag
  \\
  &
    \int_{-\infty}^{\infty}d\omega_1d\omega_2\left[\left|\psi_S\left(\omega_1,
    \omega_2\right)\right|^2 -
    \left|\psi_{AS}\left(\omega_1, \omega_2\right)\right|^2\right].
\label{eq:eq2.1.12}
\end{align}
Formula~\eqref{eq:eq2.1.12} explicitly demonstrates that the
phenomena of HOM antibunching ($P_{2c}>1/2$) and bunching
($P_{2c}<1/2$) are directly related to the symmetry properties of the
biphoton spectral function $\psi(\omega_1,\omega_2)$. Specifically, if
the spectral function is symmetric
with $\psi(\omega_1,\omega_2)=\psi(\omega_2,\omega_1)$, then $P_{2c}=0$ and photon
bunching occurs with $100\%$ probability (the perfect bunching).
In the opposite limiting case where the spectral function
is antisymmetric
and $\psi(\omega_1,\omega_2)=-\psi(\omega_2,\omega_1)$,
antibunching occurs with $P_{2c}=1$ (the perfect antibunching).

Thus, the integral that enters 
the probability expressions given by Eqs.~\eqref{eq:eq2.1.11} and~\eqref{eq:eq2.1.12}
\begin{align}
  &
    D_S=\int_{-\infty}^{\infty}d\omega_1d\omega_2\psi(\omega_1, \omega_2)\psi^*(\omega_2, \omega_1)
    \notag
  \\
  &
    =\int_{-\infty}^{\infty}d\omega_1d\omega_2\left[\left|\psi_S\left(\omega_1,
    \omega_2\right)\right|^2 -
    \left|\psi_{AS}\left(\omega_1, \omega_2\right)\right|^2\right]
\label{eq:eq2.1.13}
\end{align}
can be used as a quantifier of the degree
of symmetry with respect to frequency permutations.

This parameter,
which will be referred to as the symmetry degree,
ranges from $-1$ to $1$ and provides an alternative
way to describe the antibunching effect.
Specifically, $D_S$ is
positive (negative) when $P_{2c}<1/2$ ($P_{2c}>1/2$)  and equals
zero at $P_{2c}=1/2$.
In this framework, the biphoton spectral
function must exhibit a negative symmetry degree to observe the
antibunching effect.
For simplicity and convenience, in our subsequent analysis,
we will use the symmetry degree parameter
$D_S$ instead of $P_{2c}$.

In the next sections, we inquire more closely into
the structural conditions
on the biphoton JSA necessary to produce negative values of $D_S$
and examine how
these conditions are connected to the degree and structure of quantum
entanglement within the biphoton state.

%%%%%%%%%%%%%%%%%%%%%%%%%%%%%%%%
\subsection{Parity conditions for Hong-Ou-Mandel antibunching}
\label{subsec:parity-cond}
%%%%%%%%%%%%%%%%%%%%%%%%%%%%%%

The above-discussed idea of characterizing antibunching through the permutation symmetry of the
JSA has been previously explored, for example, in Refs.~\cite{Wang2006, Fedrizzi2009, Cromb2023}
based on certain interpretations of relation~\eqref{eq:eq2.1.13}.  From this relation,
the perfect antibunching with $D_S=-1$ may only occur when
the JSA is antisymmetric:
$\psi(\omega_1,\omega_2) = -\psi(\omega_2,\omega_1)$.
While this conclusion being
  a direct consequence of Eq.~\eqref{eq:eq2.1.13} concerns
  the limiting case of the antibunching regime,
  in this section, we perform a  detailed analysis
  of the specifically shaped SPDC-biphotons
  and demonstrate 
  that the conditions underlying
the antibunching regime with a negative degree of symmetry,
$D_S<0$, can be governed by the properties
that are not directly related to the permutation symmetry.

Our starting point is
  the special case where the JSA of the SPDC-biphoton
  state~\eqref{eq:eq2.1.1} emerging from
  the nonlinear crystal shown in Fig.~\ref{fig:fig1} can be approximated by
  Gaussian functions of the photon frequencies of the following form~\cite{Walborn2010,Wang2006} 
\begin{align}
  &
    \psi_0(\omega_1,
    \omega_2)=N_0
    \exp{\Bigl\{-\frac{(\omega_1-\Omega)^2}{2\sigma_1^2}-\frac{(\omega_2-\Omega)^2}{2\sigma_2^2}\Bigr\}}
    \notag\\
&
  \times
  e^{i(\omega_1\tau_1+\omega_2\tau_2)}
  \exp{\Bigl\{
  %\Bigr\}}
  %\exp{\Bigl\{
  -\frac{(\omega_1+\omega_2-2\Omega)^2}{2\sigma_p^2}\Bigr\}}, 
\label{eq:eq2.1.2}
\end{align}
where $N_0=(\sigma^2_p+\sigma^2_1+\sigma^2_2)^{1/4}(\pi\sigma_p\sigma_1\sigma_2)^{-1/2}$
is the
normalization constant; $\omega_1\tau_{1}$ and $\omega_2\tau_{2}$ are
the relative phase shifts of the photons due to
propagation in space; and $\sigma_p$ is the variance parameter (the pump variance) associated with the spectral width of
the pump radiation. Note that the JSA of the form \eqref{eq:eq2.1.2} can be engineered by applying Gaussian
spectral filters of bandwidths $\sigma_1$ and $\sigma_2$
to the signal and idler photons of a near-degenerate SPDC
pair, respectively.

Based on the structure of the SPDC biphoton spectrum~\eqref{eq:eq2.1.2}, we
concentrate on the modified JSA of the photon pair~\eqref{eq:eq2.1.3} taken in
the following generalized form:
\begin{align}
  &
    \psi(\omega_1,
    \omega_2)=N\varphi_{1}(\omega_1)\varphi_{2}(\omega_2)
    \notag
  \\
  &
    \times\exp{\left\{-\frac{(\omega_1+\omega_2-2\Omega)^2}{2\sigma_p^2}\right\}},
\label{eq:eq2.2.1}
\end{align}
where $\varphi_{1}(\omega)$and $\varphi_{2}(\omega)$ are some functions
representing the local spectral contributions to the JSA from the signal and idler
  photons, respectively; and $N$ is the normalization constant.  After
substituting Eq.~\eqref{eq:eq2.2.1} into Eq.~\eqref{eq:eq2.1.13}, we can express the symmetry degree
as follows
\begin{align}
  &
    D_S=N^2\int_{-\infty}^{\infty}d\tilde{\omega}_1d\tilde{\omega}_2\varphi_{12}(\tilde{\omega}_1)
    \cnj{\varphi}_{12}(\tilde{\omega}_2)
    \notag
  \\
  &
    \times\exp{\left\{-\frac{(\tilde{\omega}_1+\tilde{\omega}_2)^2}{\sigma_p^2}\right\}},
\label{eq:eq2.2.2}
\end{align}
where $\tilde{\omega}_{i}=\omega_i-\Omega$,
in terms of the spectral function
\begin{align}
  \label{eq:phi_12}
  \varphi_{12}(\omega)=\varphi_1(\omega)\cnj{\varphi}_2(\omega).
\end{align}
An important point is that,
in the large pump variance limit $\sigma_p\rightarrow \infty$,
the exponential factor that enters the spectral
distribution~\eqref{eq:eq2.2.1} approaches
unity so that the resulting two-photon state becomes separable,
whereas, in the opposite case,
where $\sigma_p\rightarrow 0$,
the factor reduces to the Dirac delta function,
$\delta(\omega_1+\omega_2-2\Omega)$,
representing the limit of maximally entangled states.
Thus, the parameter $\sigma_p$ appears to be directly related to the degree of quantum
entanglement that will be analyzed in Sec.~\ref{sec:sec3}.

Our task now is to identify the properties of the JSA~\eqref{eq:eq2.2.1} that
  govern the bunching and antibunching regimes
of the HOM interference
  determined by the sign of
  the symmetry degree~\eqref{eq:eq2.2.2}
 ($D_S>0$ and $D_S<0$, respectively).
  To this end, we begin with the case where
  $\sigma_p>0$ and treat the singular limit $\sigma_p\rightarrow0$ separately.
%
%%%%%%%%%%%%%%%%%%%%%%
\subsubsection{Non-zero pump variance
    $\sigma_p$}
\label{subsubsec:sigma-ne-0}
%%%%%%%%%%%%%%%%%%%%%%

  Our first remark concerns the well-known result that
  separable biphoton states cannot lead to antibunching with $D_S<0$ naturally arises within
  the model under consideration.
  Indeed, as discussed above, the separability of the
  biphoton state~\eqref{eq:eq2.1.3} is equivalent to the limit $\sigma_p\rightarrow\infty$. From
  Eq.~\eqref{eq:eq2.2.2} it is not difficult to obtain the limiting value of the symmetry degree

\begin{align}
  &
    \lim_{\sigma_p\to\infty}D_{S}=D_{\infty}=
    N^2\left|\int_{-\infty}^{\infty}\varphi_{12}(\tilde{\omega}) d\tilde{\omega}\right|^2\ge
    0
\label{eq:eq2.2.10}
\end{align}
that cannot be negative.

An important consequence of Eq.~\eqref{eq:eq2.2.10}
  is that, for non-zero $\sigma_p$, 
  the JSA \eqref{eq:eq2.2.1} cannot be antisymmetric:
  $\psi(\omega_1,\omega_2)\not=-\psi(\omega_2,\omega_1)$.

To see this, let us assume that $\sigma_p>0$ and
  the JSA~\eqref{eq:eq2.2.1} is antisymmetric,
$\psi(\omega_1,\omega_2)=-\psi(\omega_2,\omega_1)$, so that $D_S=-1$ (see Eq.~\eqref{eq:eq2.1.13}).
From Eq.~\eqref{eq:eq2.2.1},
this assumption is equivalent to the antisymmetry relation
$\varphi_1(\omega_1)\varphi_2(\omega_2)=-\varphi_1(\omega_2)\varphi_2(\omega_1)$
which is clearly independent of $\sigma_p$
and thus remains applicable to
the limiting case of separable states where $\sigma_p\to\infty$.
The latter leads to the conclusion that separable states
can exhibit perfect antibunching with $D_S=-1$ 
which is in sharp contradiction with
the separability condition Eq.~\eqref{eq:eq2.2.10}.

So, we have shown that $\psi(\omega_1,\omega_2)\neq -\psi(\omega_2,\omega_1)$
  at $\sigma_p>0$. This means that the biphoton with the JSA of the form~\eqref{eq:eq2.2.1} can never produce
  perfect antibunching with $D_{S}=-1$, unless the state is maximally entangled.
  Nevertheless, an important point is that the
  antibunching regime with $D_S<0$ is not prohibited at $0<\sigma_p<\infty$.
  The latter follows from the parity conditions controlling both the sign of $D_S$
  and the regime of the HOM interference. These conditions will be deduced in this section.

Our first step is to decompose
the Gaussian function
$\exp{\{-(\tilde{\omega}_1+\tilde{\omega}_2)^2/\sigma_p^2\}}$
that enters the integrand on the right hand side of Eq.~\eqref{eq:eq2.2.2}
using the Mehler formula~\cite{Bateman1953}
\begin{align}
  &
    \exp{\left\{2xy\frac{z}{1-z^2}\right\}}\exp{\left\{-(x^2+y^2)\frac{z^2}{1-z^2}\right\}}
    \notag
  \\
  &
   = \sqrt{1-z^2}\sum_{n=0}^{\infty}\frac{z^n}{2^nn!}H_n(x)H_n(y),
\label{eq:eq2.2.3}
\end{align}
where $H_n(x)$ are the Hermite polynomials.

It is rather straightforward to check that
this function can be written in the form
\begin{align}
  &
    \exp{\left\{-\frac{(\tilde{\omega}_1+\tilde{\omega}_2)^2}{\sigma_p^2}\right\}}=
    \exp{\left\{-\frac{\tilde{\omega}_1^2+\tilde{\omega}_2^2}{\sigma_p^2}q^2\right\}}
    \notag
  \\
  &
    \times\sqrt{ q}\sum_{n=0}^{\infty}\frac{(-q)^n}{2^nn!}H_n(\tilde{\omega}_1/\sigma_p)H_n(\tilde{\omega}_2/\sigma_p),
\label{eq:eq2.2.4}
\end{align}
where $q=(\sqrt{5}-1)/2$,
that immediately leads to the following representation
for the symmetry degree:
\begin{align}
  &
    D_S=N^2\sqrt{ q}\sum_{n=0}^{\infty}\frac{(-q)^n}{2^nn!}
    \notag
  \\
  &
    \times
    \left|\int_{-\infty}^{\infty}d\tilde{\omega}\varphi_{12}(\tilde{\omega})H_n(\tilde{\omega}/\sigma_p)e^{-\tilde{\omega}^2q^2/\sigma_p^2}\right|^2.
\label{eq:eq2.2.5}
\end{align}
There are two important points to note:
(a)~the well-known identity $H_n(-x)=(-1)^nH_n(x)$ implies
that $H_{2n}(x)$ are even functions, whereas
$H_{2n+1}(x)$ are odd functions;
(b)~the function $\varphi_{12}(\tilde{\omega})$ can be written as a sum,
$\varphi_{12}(\tilde{\omega})=\varphi^{(+)}_{12}(\tilde{\omega})+\varphi^{(-)}_{12}(\tilde{\omega})$,
of even and odd functions given by
\begin{align}
  &
    \varphi^{(\pm)}_{12}(\tilde{\omega})=
    \frac{1}{2}[\varphi_{12}(\tilde{\omega})\pm\varphi_{12}(-\tilde{\omega})]=\pm\varphi^{(\pm)}_{12}(-\tilde{\omega}).
\label{eq:eq2.2.6}
\end{align}

We can now express the symmetry degree~\eqref{eq:eq2.2.5}
in terms of the even and odd components of $\varphi_{12}$ as follows
\begin{align}
& D_{S} = D_{S}^{(+)} - D_{S}^{(-)},
\label{eq:eq2.2.7}
\end{align}
where
\begin{align}
  &
    D_{S}^{(+)}=N^2\sqrt{q}\sum_{n=0}^{\infty}\frac{q^{2n}}{2^{2n}(2n)!}f_{2n}^{(+)},
    \notag
  \\
  &
    D_{S}^{(-)}=N^2\sqrt{q}\sum_{n=0}^{\infty}\frac{q^{2n+1}}{2^{2n+1}(2n+1)!}f_{2n+1}^{(-)},
    \notag
  \\
  &
    f_n^{(\pm)}=\left|
    \int_{-\infty}^{\infty}d\tilde{\omega}
    \varphi_{12}^{(\pm)}(\tilde{\omega})e^{-\tilde{\omega}^2q^2/\sigma_p^2}H_{n}\left(\frac{\tilde{\omega}}{\sigma_p}\right)
    \right|^2,
\label{eq:eq2.2.8}
\end{align}
where $f_{2n}^{(-)}=f_{2n+1}^{(+)}=0$ due to the
parity properties of the Hermite polynomials and
$\varphi_{12}^{(\pm)}(\tilde{\omega})$.

Equation~\eqref{eq:eq2.2.7} clearly shows
that $D_S$ can be governed by the parity of the function
$\varphi_{12}(\tilde{\omega})$:
the positive (negative) part of $D_S$ is determined
by the even (odd) component of $\varphi_{12}$,
$\varphi_{12}^{(+)}$ ($\varphi_{12}^{(-)}$).
In particular, for $0<\sigma_p<\infty$,
the regime of bunching (antibunching) is always achieved
when
the spectral
function $\varphi_{12}(\omega)=\varphi_1(\omega)\cnj{\varphi}_2(\omega)$
meets the even (odd) parity condition given by
\begin{align}
  &
    \varphi_{12}(\omega-\Omega)=\pm\varphi_{12}(\Omega-\omega).
\label{eq:eq2.2.9}
\end{align}
For separable states, the symmetry degree vanishes provided that
  $\varphi_{12}(\tilde{\omega})=\varphi_{12}^{(-)}(\tilde{\omega})$ (see Eq.~\eqref{eq:eq2.2.10}),
  whereas all other spectral functions will give the positive value of $D_S$.
More generally, in the case where both $\varphi_{12}^{(+)}(\tilde{\omega})$ and
  $\varphi_{12}^{(-)}(\tilde{\omega})$ are non-vanishing, the sign of $D_S$ is determined by the
  interplay between the even and odd contributions. Thus, the odd parity of the spectral function
  $\varphi_{12}(\tilde{\omega})$ is a strong requirement
  eliminating the even contribution $D_S^{(+)}$ to ensure
  the negative sign of $D_S$, $D_S=-D_S^{(-)}<0$.
  It should be stressed that,
  since the JSA cannot be antisymmetric at $\sigma_p>0$,
  in this case,
  the symmetric contribution,
  $\int d\omega_1d\omega_2|\psi_S(\omega_1,\omega_2)|^2$, to $D_S$ (see Eq.~\eqref{eq:eq2.1.12})
  cannot be zero even if $D_S^{(+)}=0$.

The above discussion emphasizes the difference between
  the parity conditions~\eqref{eq:eq2.2.9} and the symmetry
  conditions that represent two different mechanisms to control the value of the symmetry degree
  (and the regime of the HOM interference).
  The former can be regarded as strong constraints imposed on the spectral function to
  fix the sign of the symmetry degree,
  whereas the latter ensures the extreme values of $D_S$.

%%%%%%%%%%%%%%%%%
\subsubsection{Limit of vanishing pump variance
    $\sigma_p$: $\sigma_p\to 0$}
\label{subsubsec:sigma-eq-0}
%%%%%%%%%%%%%%%

For the sake of brevity, in this subsection,
  we just summarize
  the main points leaving aside technical details that can be found in Appendix~\ref{append:sp=0}.

In the limit
  $\sigma_p\rightarrow0$,
  the entangling factor of the JSA~\eqref{eq:eq2.2.1}
  becomes the $\delta$ function: $\delta(\tilde\omega_2+\tilde\omega_1)$.

At $\tilde\omega_2=-\tilde\omega_1$, the JSA (anti)symmetry conditions
  $\psi(\tilde\omega_1,-\tilde\omega_1)=\pm\psi(-\tilde\omega_1,\tilde\omega_1)$, applied to the
  spectral distribution~\eqref{eq:eq2.2.1}, can be written in the form of the parity conditions
\begin{align}
  &\varphi_1(\omega-\Omega)\varphi_2(\Omega-\omega)
 =
    \pm\varphi_1(\Omega-\omega)\varphi_2(\omega-\Omega).
\label{eq:perfect_antibunch}
\end{align}
An important point is that Eq.~\eqref{eq:perfect_antibunch} is not equivalent
  to the conditions~\eqref{eq:eq2.2.9}.
  So, the parity conditions~\eqref{eq:eq2.2.9} do not guarantee
  perfect (anti)bunching with $D_{\pm}=\pm 1$ without imposing
  additional constraints.
It can be shown that such perfect (anti)bunching constraints require the
  magnitude of either $\varphi_{1}(\tilde{\omega})$ or $\varphi_{2}(\tilde{\omega})$ to be an even
  function of frequency:
\begin{align}
&
|\varphi_{i}(-\tilde{\omega})|^2=|\varphi_{i}(\tilde{\omega})|^2.
\label{eq:extra_cond}
\end{align}
Thus, Eq.~\eqref{eq:extra_cond} combined with the odd-parity condition leads to
Eq.~\eqref{eq:perfect_antibunch}. However, the converse is not true. In other words,
Eq.~\eqref{eq:perfect_antibunch} can generally be satisfied even if Eq.~\eqref{eq:eq2.2.9} does not
hold.

      Our concluding remark concerns the case
      represented by the modulated biphotons studied in the next section.
This is the case,
where  each of the functions
    $\varphi_{1}(\tilde{\omega})$ and
    $\varphi_{2}(\tilde{\omega})$
    satisfy either the odd or even parity condition:
    $\varphi_{i}(-\tilde{\omega})=\pm\varphi_{i}(\tilde{\omega})$,
    so that Eqs.~\eqref{eq:eq2.2.9}
    and~\eqref{eq:perfect_antibunch} are both satisfied.
    By contrast,
    for distinguishable signal and idler photons
    with $\varphi_{1}(\tilde{\omega})\ne\varphi_{2}(\tilde{\omega})$,
the permutation symmetry of the JSA~\eqref{eq:eq2.2.1}
is broken provided that $\sigma_p>0$.
However, even for such photons,
the limit of vanishing pump variance
will produce the perfect (anti)bunching along with maximal entanglement.
  This comes as no surprise since the interference of a highly entangled
  two-photon state is not merely the interference of two separate photons~\cite{Shih1998,Strekalov:pra:1998}.
  In Sec.~\ref{sec:sec3} we shall illustrate this situation.  

%%%%%%%%%%%%%%%%%%%%%%%%%%%%%%%% 
\subsection{Even parity breaking in joint spectral amplitudes}
\label{subsec:parity-breaking}
%%%%%%%%%%%%%%%%%%%%%%%%%%%%%%

For the JSA of the standard SPDC state~\eqref{eq:eq2.1.2},
the \textcolor{magenta}{local} spectral functions that enter the generalized
representation~\eqref{eq:eq2.2.1} are
$\varphi_{1,2}(\omega)=\exp\{i\omega\tau_{1,2}-(\omega-\Omega)^2/(2\sigma_{1,2}^2)\}$,
where $\tau_{1}$ and $\tau_{2}$ determine the phase accumulated
  by the photon due to propagation in space (see Eq.~\eqref{eq:eq2.1.2}).
At $\tau_1=\tau_2$,
when the distances that photons travel from a nonlinear crystal to BS are equal (see Fig.~\ref{fig:fig1}),
this JSA satisfies the even-parity condition:
$\varphi_{12}(\omega-\Omega)=\varphi_{12}(\Omega-\omega)$.
As a result, the symmetry degree $D_S$ is positive, and
no antibunching may occur.

At $\tau_1\neq \tau_2$,
the even parity appears to be broken
because
the difference in the phase delays
introduces an additional phase factor
$\exp{\{i\omega (\tau_1-\tau_2)\}}$
into the spectral function
$\varphi_{12}(\omega)$ (see Eq.~\eqref{eq:phi_12}).

In this case, the symmetry degree
can be computed by
performing Gaussian integrals.
The resulting expression
\begin{align}
  &
    D_S=
    \frac{2\sigma_1\sigma_2\sqrt{\sigma_1^2+\sigma_2^2+\sigma_p^2}}{%
    \sqrt{\sigma_1^2+\sigma_2^2}\sqrt{\sigma_p^2(\sigma_1^2+\sigma_2^2)+4\sigma_1^2\sigma_2^2}
    }
    \notag
  \\
  &
    \times\exp{\left\{-\Delta\tau^2\frac{\sigma_1^2\sigma_2^2}{\sigma_1^2+\sigma_2^2}\right\}}\ge0,
\label{eq:eq2.2.11}
\end{align} 
where $\Delta\tau=\tau_2-\tau_1$,
clearly implies that asymmetry in the phase delays
does not lead to HOM antibunching.
From Eq.~\eqref{eq:eq2.2.11},
the symmetry degree monotonically decreases with
the magnitude of the time delay difference, $|\Delta\tau|$,
approaching zero in the limit $|\Delta\tau|\rightarrow\infty$.

Since the standard biphoton state cannot exhibit antibunching,
in order to observe this effect, the SPDC biphoton JSA must be appropriately
modified.
In Sec.~\ref{sec:sec4}, we will detail the
engineering of modulated SPDC biphotons
by incorporating a Mach–Zehnder
interferometer (MZI)  into one arm of
a standard HOM interferometer~\cite{Shih1998,Shih2020}. 
Our result is that
the modified local spectral function of
the modulated JSA is given by
\begin{align}
  &
    \varphi_2(\omega_2)
    =\exp{\left\{-\frac{(\omega_2-\Omega)^2}{2\sigma_2^2}\right\}}e^{i\omega_2\tau_2}
    \varphi_c(\omega_2),
    \notag
  \\
  &
    \varphi_c(\omega_2) =
    \cos{\Bigl(\frac{\omega_2\Delta L}{2c}\Bigr)},
\label{eq:eq2.2.12}
\end{align}
where
$\Delta L$ is the path-length difference in the arms of the MZI,
and
$c$ is the speed of light. 
Note that MZI-modulation can also be tuned so as
to have the cosine-factor
$\varphi_c(\omega_2) $
replaced with the sine-factor
\begin{align}
  \label{eq:phi_s}
  \varphi_s(\omega_2) =\sin{\left(\frac{\omega_2\Delta L}{2c}\right)}.
\end{align}

Parity properties of the cosine-factor
$\varphi_c(\omega)=\cos(\beta\omega)$
depend on the value of the coefficient
$\beta=\Delta L/(2c)$.
The cosine-factors that meet
the parity condition
$\varphi_c^{(\pm)}(\omega-\Omega)=\pm \varphi_c^{(\pm)}(\Omega-\omega)$
can be written in the form
\begin{align}
  &
    \varphi_c^{(-)}\left(\omega\right) =
    \cos{\frac{\omega\pi(n+1/2)}{\Omega}},
  %   \notag
  % \\
  % &
\;
  \varphi_c^{(+)}\left(\omega\right) = \cos{\frac{\omega\pi n}{\Omega}},
\label{eq:eq2.2.13}
\end{align}
where
$n\in\mathbb{N}$ is an integer.
The corresponding result for the sine-factors,
$\varphi_s^{(\pm)}(\omega-\Omega)=\pm \varphi_s^{(\pm)}(\Omega-\omega)$,
reads
\begin{align}
  &
    \varphi_s^{(-)}\left(\omega\right) = \sin{\frac{\omega\pi n}{\Omega}},
  %       \notag\\
  %       &
\;
    \varphi_s^{(+)}\left(\omega\right) = \sin{\frac{\omega\pi(n+1/2)}{\Omega}}.
\label{eq:eq2.2.14}
\end{align}
Clearly,
at $\tau_1=\tau_2$ and $\varphi_c=\varphi_c^{(-)}$
($\varphi_c=\varphi_c^{(+)}$),
the modified JSA with the spectral function \eqref{eq:eq2.2.12}
represents the biphoton state that shows up in
the antibunching (bunching) regime, including the perfect antibunching (bunching).
Thus, we can dynamically
switch between bunching and antibunching
with either $\Delta L/(2c)=n\pi/ \Omega$
or $\Delta L/(2c)=(n+1/2)\pi/ \Omega$
by controlling the length of MZI arms.

Since such switching involves
changing
the sign of the symmetry degree,
it cannot occur without affecting the degree of entanglement of
the modulated biphoton.
In order to study the effects related to the biphoton entanglement,
the parameters of the biphoton spectrum should be linked
to a well-defined quantitative measure of entanglement.
In subsequent sections, this issue will be examined in detail.

In conclusion of this section, it should be emphasized
  that one of our key results is the parity conditions~\eqref{eq:eq2.2.9}.
  As opposed to the perfect (anti)bunching
  conditions~\eqref{eq:perfect_antibunch}
arising from the well-known (anti)symmetry conditions
(see, e.g., formulas (38) and (59) in Ref.~\cite{Wang2006}) in the limit
of vanishing pump variance,
  the conditions~\eqref{eq:eq2.2.9} provide a means to control the sign of the symmetry
  degree that governs regimes of the HOM interference at an arbitrary pump variance.

%%%%%%%%%%%%%%%%%%%%%%%%%%%
\section{Entanglement of biphoton states}
\label{sec:sec3}
%%%%%%%%%%%%%%%%%%%%%%%%%%%

As discussed in Sec.~\ref{subsec:parity-breaking},
the biphoton state~\eqref{eq:eq2.1.2} produced via
the SPDC process,
which we refer to as the standard biphoton state, cannot demonstrate
the antibunching effect.
Nevertheless,
in order to illustrate our general approach used
to characterize the entanglement degree
of biphoton states and its connection with the symmetry degree,
we begin with the case of Gaussian SPDC-biphotons.
One of the objectives of this section is, thus,
to lay a foundation for the characterization of the entanglement
properties of modulated biphoton states
that will be studied in the next section.

The standard method
for analyzing the entanglement properties in pure quantum states of
bipartite continuous-variable systems is
the Schmidt decomposition technique,
where the state~\eqref{eq:eq2.1.3} is expanded
in terms of the Schmidt modes
representing the eigenstates of
the reduced density matrices,
$\hat{\rho}_1=\Tr_2\ket{\Psi}\bra{\Psi}$
and $\hat{\rho}_2=\Tr_1\ket{\Psi}\bra{\Psi}$.
This expansion can be written in
the following form~\cite{Law2000, Parker2000,
  Fedorov2014}:
\begin{align}
\left|\Psi\right\rangle =  \sum_{n=0}^{\infty}\sqrt{\lambda_n}\left|\phi_n\right\rangle_{1}\left|\vartheta_n\right\rangle_{2},
\label{eq:eq3.1.1}
\end{align}
where ${\lambda_n}\neq 0$ are the Schmidt eigenvalues
that meet the normalization condition $\sum_{n=0}^{\infty} \lambda_n=1$;
${\left|\phi_n\right\rangle}$ and ${\left|\vartheta_n\right\rangle}$
are the Schmidt modes that are subject to
the orthogonality conditions:
$\langle\phi_n|\phi_m\rangle=\langle\vartheta_n|\vartheta_m\rangle=\delta_{nm}$.
One of the widely used quantitative measures of entanglement
is the Schmidt number
\begin{align}
  \label{eq:Schidt-num}
  K= \Bigl\{\sum_{n=0}^{\infty}\lambda_n^2\Bigr\}^{-1},
\end{align}
which is inversely proportional to the purity
of the reduced density matrices,
$K=1/\Tr\hat{\rho}_i^2$.
At $K=1$, the state is separable,
whereas $\ket{\Psi}$ is an entangled state
provided that $K> 1$.
So, the larger the Schmidt number,
the more entangled the state.
For continuous-variable systems, the value of $K$
is not bounded from above.

To apply the Schmidt decomposition for the standard SPDC state,
we introduce the two variables:
$x=(\omega_1-\Omega)/s_1$ and $y=(\Omega-\omega_2)/s_2$,
where $s_1$ and $s_2$ are the parameters to be determined,
and recast the Mehler formula~\eqref{eq:eq2.2.3}
into the following suitably modified form:
\begin{align}
  &
    N_0
    \sqrt{\pi}\sqrt{1-z^2}\sum_{n=0}^{\infty}z^ne^{i(xs_1+\Omega)\tau_1}\psi_n(x)e^{i(-ys_1+\Omega)\tau_1}\psi_n(y)
    \notag
  \\
  &
    =N_0\exp{\left\{2xy\frac{z}{1-z^2}-\frac{(x^2+y^2)}{2}\frac{1+z^2}{1-z^2}\right\}}
    \notag
  \\
  &
    \times
    e^{i(xs_1+\Omega)\tau_1}e^{i(-ys_1+\Omega)\tau_1},
\label{eq:eq3.1.2}
\end{align}
where $\psi_n(x)=\left(\sqrt{\pi}n!2^n\right)^{-1/2}H_n(x)\exp(-x^2/2)$
are the quantum oscillator eigenfunctions
and $z$ is the parameter
whose absolute value cannot exceed unity: $|z|\le 1$.

There are three parameters,
$s_{1}$, $s_{2}$, and $z$, that enter the modified Mehler formula~\eqref{eq:eq3.1.2}
and can be found by assuming that the right side of  Eq.~\eqref{eq:eq3.1.2}
equals the JSA~\eqref{eq:eq2.1.2}.
After rather straightforward and lengthy algebraic manipulations, we have
\begin{align}
  &
    s_1=\frac{\sigma_1\sqrt{\sigma_p}(\sigma_2^2+\sigma_p^2)^{1/4}}{(\sigma_1^2+\sigma_p^2)^{1/4}(\sigma_1^2+\sigma_2^2+\sigma_p^2)^{1/4}},
    \notag
  \\
  &
    s_2=s_1
    \frac{\sigma_2}{\sigma_1}\sqrt\frac{\sigma_1^2+\sigma_p^2}{\sigma_2^2+\sigma_p^2},
    \notag
  \\
  &
    z=\sqrt{\frac{s_1^2(\sigma_1^2+\sigma_p^2)-\sigma_1^2\sigma_p^2}{s_1^2(\sigma_1^2+\sigma_p^2)+\sigma_1^2\sigma_p^2}}.
\label{eq:eq3.1.5}
\end{align}

Since the left-hand side of
Eq.~\eqref{eq:eq3.1.2} gives the JSA of
the quantum state,
we obtain the Schmidt modes 
\begin{align}
  &
    \left|\phi_n\right\rangle_{1}\notag =
    \frac{1}{\sqrt{s_1}}\int_{-\infty}^{\infty}d\omega_1\psi_n\left(\frac{\omega_1-\Omega}{s_1}\right)e^{i\omega_1\tau_1}\hat{a}_1^{\dagger}(\omega_1)|0\rangle_1,
    \notag
  \\
  &
    \left|\vartheta_n\right\rangle_{2} =
    \frac{(-1)^n}{\sqrt{s_2}}\int_{-\infty}^{\infty}d\omega_2\psi_n\left(\frac{\omega_2-\Omega}{s_2}\right)e^{i\omega_2\tau_2}\hat{a}_2^{\dagger}(\omega_2)|0\rangle_2,
\label{eq:eq3.1.3}
\end{align}
along with the normalization coefficient $N_0=1/\sqrt{s_1s_2\pi}$
and  the Schmidt eigenvalues $\lambda_n=(1-z^2)z^{2n}$.
The latter can now be combined with
relations~\eqref{eq:Schidt-num} and~\eqref{eq:eq3.1.5} 
 to deduce the expression for the Schmidt number
\begin{align}
  &
    K = \frac{1+z^2}{1-z^2} =
    \sqrt{1+\frac{\sigma_1^2\sigma_2^2}{\sigma_p^2(\sigma_1^2+\sigma_2^2+\sigma_p^2)}}.
\label{eq:eq3.1.4}
\end{align}
It comes as no surprise that, by contrast to
the symmetry degree~\eqref{eq:eq2.2.11},
the Schmidt number~\eqref{eq:eq3.1.4} is independent of $\Delta\tau$.
Indeed, the spatial phase shift is generated by
a local unitary transformation of the photonic modes,
and such a transformation cannot influence the entanglement degree.

We can now use Eqs.~\eqref{eq:eq2.2.11}
and~\eqref{eq:eq3.1.4}
to analyze how the symmetry degree
$D_S$ of the SPDC state depends on the Schmidt number.

For this purpose,
we shall assume that,
in real-world experiments, it is easier to
tune the pump width, $\sigma_p$, rather
than the variance parameters $\sigma_1$ and $\sigma_2$.
This is why it is reasonable to
express the parameter $\sigma_p$ in terms of the Schmidt number.
By solving Eq.~\eqref{eq:eq3.1.4} for $\sigma_p$, we have
\begin{align}
  &
    \sigma_p^2=
    \sqrt{\frac{(\sigma_1^2+\sigma_2^2)^2}{4}+\frac{\sigma_1^2\sigma_2^2}{K^2-1}}-\frac{\sigma_1^2+\sigma_2^2}{2}.
\label{eq:eq3.1.7}
\end{align}
In agreement with our discussion in Sec.~\ref{subsec:parity-cond},
this formula implies that,
in the limiting case of separable states
with $K\to 1$,
the parameter $\sigma_p$ becomes infinitely large,
$\sigma_p\to\infty$.
By contrast, $\sigma_p$ vanishes, provided
the Schmidt number increases indefinitely,
approaching the limit of an extremely entangled state. 

\begin{figure}
\centering 
        \includegraphics[width=1\linewidth]{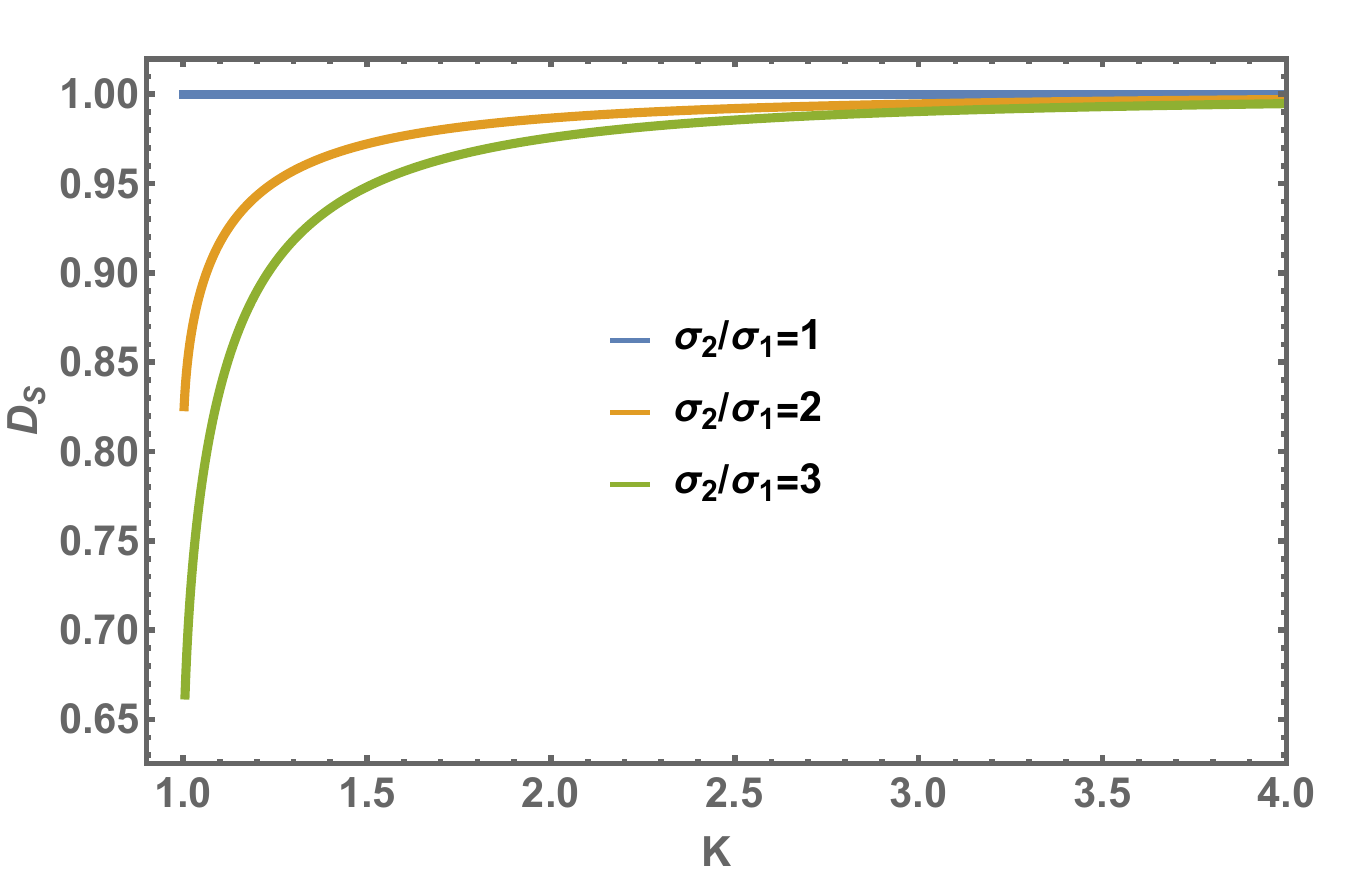}
\caption{Dependence of the symmetry degree, $D_S$, of SPDC biphoton
  joint spectral
  amplitude with the Schmidt modes~\eqref{eq:eq3.1.3} on
  the Schmidt number $K$ computed from
  Eqs.~\eqref{eq:eq2.2.11} and~\eqref{eq:eq3.1.4}
at different values of the variance ratio $\sigma_2/\sigma_1$
  with $\Delta \tau = 0$. 
  %fixed $\sigma_1, \sigma_2$ ($\Omega=2\pi\times844.5$ THz, $\sigma_1=2\pi\times10$ THz).
}
\label{fig:fig2}
\end{figure}

By substituting Eq.~\eqref{eq:eq3.1.7}
into relation~\eqref{eq:eq2.2.11},
we can compute the symmetry degree, $D_S$,
as a function of the Schmidt number, $K$,
assuming that the parameters
$\sigma_1$ and $\sigma_2$ are fixed
and $\sigma_p$ varies with $K$.
The results are presented in Fig.~\ref{fig:fig2}.

Note that the symmetry degree~\eqref{eq:eq2.2.11}
depends on the three dimensionless parameters:
$\sigma_p^2/(\sigma_1\sigma_2)$,
$\sigma_1/\sigma_2+\sigma_2/\sigma_1$
and $\Delta \tau\sqrt{\sigma_1\sigma_2}$,
whereas Eq.~\eqref{eq:eq3.1.7}
implies that the ratio $\sigma_p^2/(\sigma_1\sigma_2)$ is a function
of the Schmidt number, $K$, and the symmetrized sum of ratios
$\sigma_1/\sigma_2+\sigma_2/\sigma_1$. Thus, in the case $\Delta\tau=0$, $D_S$ is governed by $K$ and
the ratio $r=\sigma_2/\sigma_1$ independently on the absolute value of $\sigma_1$ or $\sigma_2$.
Figure~\ref{fig:fig2} shows the curves evaluated at $\Delta\tau=0$. Clearly, the only effect  of non-vanishing $\Delta\tau$
is reduction of the magnitude of $D_S$ described by the last
exponential factor on the right-hand side of Eq.~\eqref{eq:eq2.2.11}.

The JSA is symmetric with respect to frequency permutations,
$\psi(\omega_1,\omega_2)=\psi(\omega_2,\omega_1)$, provided that $\Delta\tau=0$ and
$\sigma_1=\sigma_2$ ($r=1$).  From Eq.~\eqref{eq:eq2.2.11}, at $\sigma_1=\sigma_2\equiv \sigma$,
$D_S$ equals $\exp[-(\Delta\tau\sigma)^2/2]$ and thus is independent of $\sigma_p$.  The Schmidt
number, in its turn, is governed by the ratio $\sigma_p/\sigma$:
$K=\sqrt{1+(\sigma_p/\sigma)^{-2}/(2+(\sigma_p/\sigma)^{2})}$.  Given the value of $\sigma$, it
turned out that the degree of entanglement has no effect on the symmetry degree $D_S$.
This is why
there is no entanglement-antibunching correlation
in the symmetric case
previously considered in Refs.~\cite{Fedorov2009, Zielnicki2018}. 

So, as expected for the biphoton with symmetric JSA, we have the regime of perfect
bunching with $D_S=1$ (see the blue curve in Fig.~\ref{fig:fig2}).  At $\sigma_1\neq\sigma_2$
($r\ne 1$), the permutation symmetry is broken and $D_S$ remains positive due to the even parity
condition.  Referring to Fig.~\ref{fig:fig2}, in this case, the symmetry degree increases with the
Schmidt number approaching unity in the limit of extreme entanglement, because
the state~\eqref{eq:eq2.1.2} meets the perfect
bunching condition~\eqref{eq:perfect_antibunch}.
This is exactly the case
discussed at the end of Sec.~\ref{subsec:parity-breaking}
where the perfect bunching is the limiting regime at $\sigma_p\to 0$
for the JSA which is not symmetric
with respect to frequency permutations at $\sigma_p>0$ .
Note that the curves are invariant under the transformation where the ratio $r$ is
replaced with its inverse $1/r$.

It can also be seen from Fig.~\ref{fig:fig2} that the biphoton states with a Schmidt number of about $4$
exhibit a symmetry degree in the immediate vicinity of unity.
Note that this result is general, since it is independent of the absolute value $\sigma_1,~\sigma_2$. 
The latter renders the task of the quantitative assessment of the Schmidt number  
from the values of $D_S$ for such highly entangled states challenging.
Thus, the above estimation for the Schmidt number can be considered
as a characterization of the sensitivity threshold for positive symmetry-driven
responses in such biphoton states.

%%%%%%%%%%%%%%%%%%%%%%%%%%%
\section{Modulated  biphoton states}
\label{sec:sec4}
%%%%%%%%%%%%%%%%%%%%%%%%%%

%%%%%%%%%%%%%%%%%%%%%%%%
\subsection{Engineering}
\label{subsec:generation}
%%%%%%%%%%%%%%%%%%%%%%%%%%

As we have discussed in Sec.~\ref{subsec:parity-breaking},
achieving the
extreme antibunching regime requires the biphoton spectrum to satisfy
the specific odd parity condition~\eqref{eq:eq2.2.9}.
It was shown that this condition can be met by modulating the
standard SPDC biphoton JSA using either cosine
or sine modulating functions of
the forms given by Eqs.~\eqref{eq:eq2.2.12} and~\eqref{eq:phi_s}, respectively.

\begin{figure}[h]
\includegraphics[width=1\linewidth]{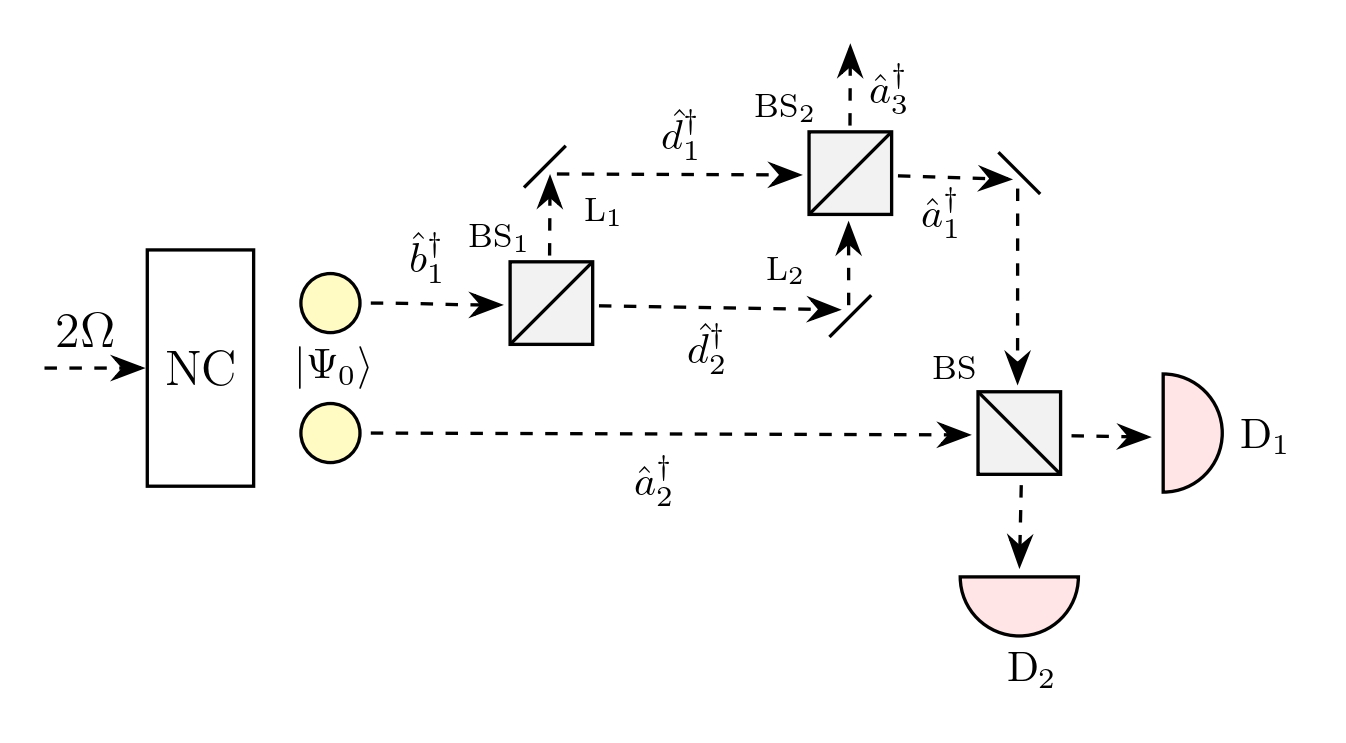}
% Here is how to import EPS art
\caption{Conceptual scheme of optical setup to modulate biphoton states.}
\label{fig:fig3} 
\end{figure}

Such modulation can be realized using
the optical setup developed in Refs.~\cite{Shih1998,Shih2020}. 
This setup is shown in Fig.~\ref{fig:fig3}
and can be regarded as an implementation of
the general scheme presented in Fig.~\ref{fig:fig1}
with the MZI incorporated into one of the arms,
thereby enabling the desired spectral
modulation through controlled interference.

In this section, we begin with a brief discussion of the process of the biphoton spectrum
modulation via the setup presented in Fig.~\ref{fig:fig3}.
The initial SPDC state after the nonlinear crystal is given by
\begin{align}
  &
    \ket{\Psi_0} =
    \notag
  \\
  &
    \int_{-\infty}^{\infty}d\omega_1d\omega_2 \psi_0\left(\omega_1,
    \omega_2\right)\hat{b}^{\dagger}_{1}
    (\omega_1)\hat{a}^{\dagger}_{2}(\omega_2)\ket{0},
\label{eq:eq4.1.1}
\end{align}
where
$\left|0\right\rangle=\left|0\right\rangle_{1}\otimes\left|0\right\rangle_{2}\otimes\left|0\right\rangle_{3}$
is the vacuum state.
The first beam splitter of the MZI, BS$_1$,
transforms the creation operator
of the first spatial mode,
$b_1^\dagger(\omega_1)$
into a superposition of
the modes,
$d_1^\dagger(\omega_1)$ and $d_2^\dagger(\omega_1)$,
propagating through the two arms
connected to the output ports of BS$_1$:
$b_1^\dagger(\omega_1)\rightarrow(d_2^\dagger(\omega_1)
e^{i\omega_1L_2/c}+id_1^\dagger(\omega_1)
e^{i\omega_1L_1/c})/\sqrt{2}$,
where $L_{1}$ and $L_{2}$ are the path lengths of
the corresponding MZI arms.
After the second beam splitter, BS$_2$, we have:
$d_1^\dagger(\omega_1)\rightarrow(a_1^\dagger(\omega_1)+ia_3^\dagger(\omega_1))/\sqrt{2}$
and
$d_2^\dagger(\omega_1)\rightarrow(a_3^\dagger(\omega_1)+ia_1^\dagger(\omega_1))/\sqrt{2}$,
where $a_1^\dagger(\omega_1)$ and
$a_3^\dagger(\omega_1)$ are the creation operators of the MZI output modes.
So, the resulting biphoton state transformed by the MZI
can be written in the form:
\begin{align}
  &\left|\Psi\right\rangle =
    \notag
  \\
  &
    i\int_{-\infty}^{\infty}d\omega_1d\omega_2 \psi_0(\omega_1, \omega_2)e^{i\omega_1L_S/(2c)}
    \Bigl[\hat{a}^{\dagger}_{1}(\omega_1)\cos{(\beta\omega_1)}
    \notag
  \\
  &+\hat{a}^{\dagger}_{3}(\omega_1)\sin{(\beta\omega_1)}
    \Bigr]
    \hat{a}^{\dagger}_{2}\left(\omega_2\right)\left|0\right\rangle,\notag\\
    &\beta=\frac{\Delta L}{2c},
\label{eq:eq4.1.2}
\end{align}
where $L_S=L_1+L_2$ and $\Delta L=L_2-L_1$.  This formula implies that we can obtain either a
cosine- or sine-modulated biphoton state by choosing a suitable output port of BS$_2$ and
performing the corresponding post-selection procedure. The efficiency of the
  post-selection procedure is dictated by the probabilities, $P_{\ind{cos}}$ and $P_{\ind{sin}}$, for the photon
  having propagated through the MZI to be found in the first (cosine-state) or third (sine-state)
  spatial mode, respectively.
  By performing the Gaussian integral
  $\int_{-\infty}^{\infty}d\omega_1d\omega_2 |\psi_0(\omega_1, \omega_2)|^2\cos(2\beta\omega_1)$,
  these probabilities can be found in the following explicit form:
\begin{align}
  &
    P_{\ind{cos}}=\frac{1+\exp\{{-\beta^2\alpha^2\}}\cos{(2\beta\Omega)}}{2},
    \notag
  \\
  &
    P_{\ind{sin}}=\frac{1-\exp{\{-\beta^2\alpha^2\}}\cos{(2\beta\Omega)}}{2},
    \notag
  \\
  &
    \alpha^2=\frac{\sigma_1^2(\sigma_2^2+\sigma_p^2)}{\sigma_1^2+\sigma_2^2+\sigma_p^2}.
\label{eq:P_cond}
\end{align}
Note that the state~\eqref{eq:eq4.1.2} can also be produced using
the optical scheme depicted in Fig.~\ref{fig:fig3} with
the MZI replaced with a Michelson interferometer.

%%%%%%%%%%%%%%%%%%%
\subsection{Symmetry degree}
\label{subsec:symmdeg}
%%%%%%%%%%%%%%%%%%%

In our subsequent analysis, we restrict ourselves to the biphoton state with the
cosine-modulated JSA
(the sine-modulated state is treated in a similar way):
\begin{align}
  &
    \psi(\omega_1, \omega_2)=N
    \exp{\left\{-\frac{(\omega_1-\Omega)^2}{2\sigma_1^2}-\frac{(\omega_2-\Omega)^2}{2\sigma_2^2}\right\}}
    \notag
  \\
  &
    \times
    e^{i(\omega_1\tau_1+\omega_2\tau_2)}\cos{\beta\omega_1}
    \exp{\left\{-\frac{(\omega_1+\omega_2-2\Omega)^2}{2\sigma_p^2}\right\}},
\label{eq:eq4.1.3}
\end{align}
where $\beta$ is the tunable time-delay parameter described in
  Eq.~\eqref{eq:eq4.1.2},
  and $N=N_0P_{\ind{cos}}^{-1/2}$ is the normalization constant.

The symmetry degree of the state in Eq.~\eqref{eq:eq4.1.3} can be
evaluated analytically as follows:
\begin{align}
  &
    D_S = D_S^{(0)}\bigl\{
    e^{-\beta^2\gamma^2\sigma_p^2}
    \cos(2\beta\Omega)+e^{-\beta^2\xi^2}\cosh(2\beta\Delta\tau\xi^2)
    \bigr\}
    \notag
  \\
  &
    \times
    \bigl[
    1+e^{-\beta^2\alpha^2}\cos{(2\beta\Omega)}
    \bigr]^{-1}
\label{eq:eq4.1.4}
\end{align}
where
$D_S^{(0)}$ and $\alpha$ are given by Eq.~\eqref{eq:eq2.2.11} and Eq.~\eqref{eq:P_cond},
respectively;
$\gamma^2=\sigma_1^2\sigma_2^2/[(\sigma_1^2+\sigma_2^2)\sigma_p^2+4\sigma_1^2\sigma_2^2]$,
and $\xi^2=\sigma_1^2\sigma_2^2/(\sigma_1^2+\sigma_2^2)$.

From Eq.~\eqref{eq:eq4.1.4},
by assuming the ideal case, where $\Delta\tau=0$
and
$\sigma_p\rightarrow0$,
and the limit of small modulation time delay
with $\beta^2\xi^2\ll1$,
the symmetry degree
can be put into the simplified form:
\begin{align}
  &D_S|_{\beta^2\xi^2\ll1} \approx
    \notag
  \\
  &
    1+\frac{2\beta^2\xi^2\cos{(2\beta\Omega)}}{1+\cos{(2\beta\Omega)}-\beta^2\xi^2\cos{(2\beta\Omega)}}.
\label{eq:eq4.1.5}
\end{align}
This expression exhibits sharp resonances at the points
$\beta=\beta_n = \pi(2n + 1)/(2\Omega)$, where the symmetry degree reaches its
minimum value, $D_S = -1$. Away from these resonance points, the
second term in Eq.~\eqref{eq:eq4.1.5} becomes negligible,
so that
$D_S$ is close to unity.
Since $\beta$ is the continuously varying parameter,
the $\beta$-dependence of $D_S$ at $\beta^2\xi^2 \ll 1$
features pronounced narrow dips to negative values.
In the opposite regime of large time delays with
$\beta^2\xi^2 \gg 1$
the symmetry degree asymptotically approaches
$D_S \approx \cos{(2\beta\Omega)}$. This behavior indicates that the
resonance dips will progressively broaden as the time delay parameter
$\beta$ increases (the case of large time delays with $\sigma_1=\sigma_2$ and $\sigma_p=0$ was previously analyzed in Ref.~\cite{Wang2006}).

\begin{figure*}
\centering
      
    \begin{subfigure}[b]{0.3\textwidth}
        \centering
        \includegraphics[width=\linewidth]{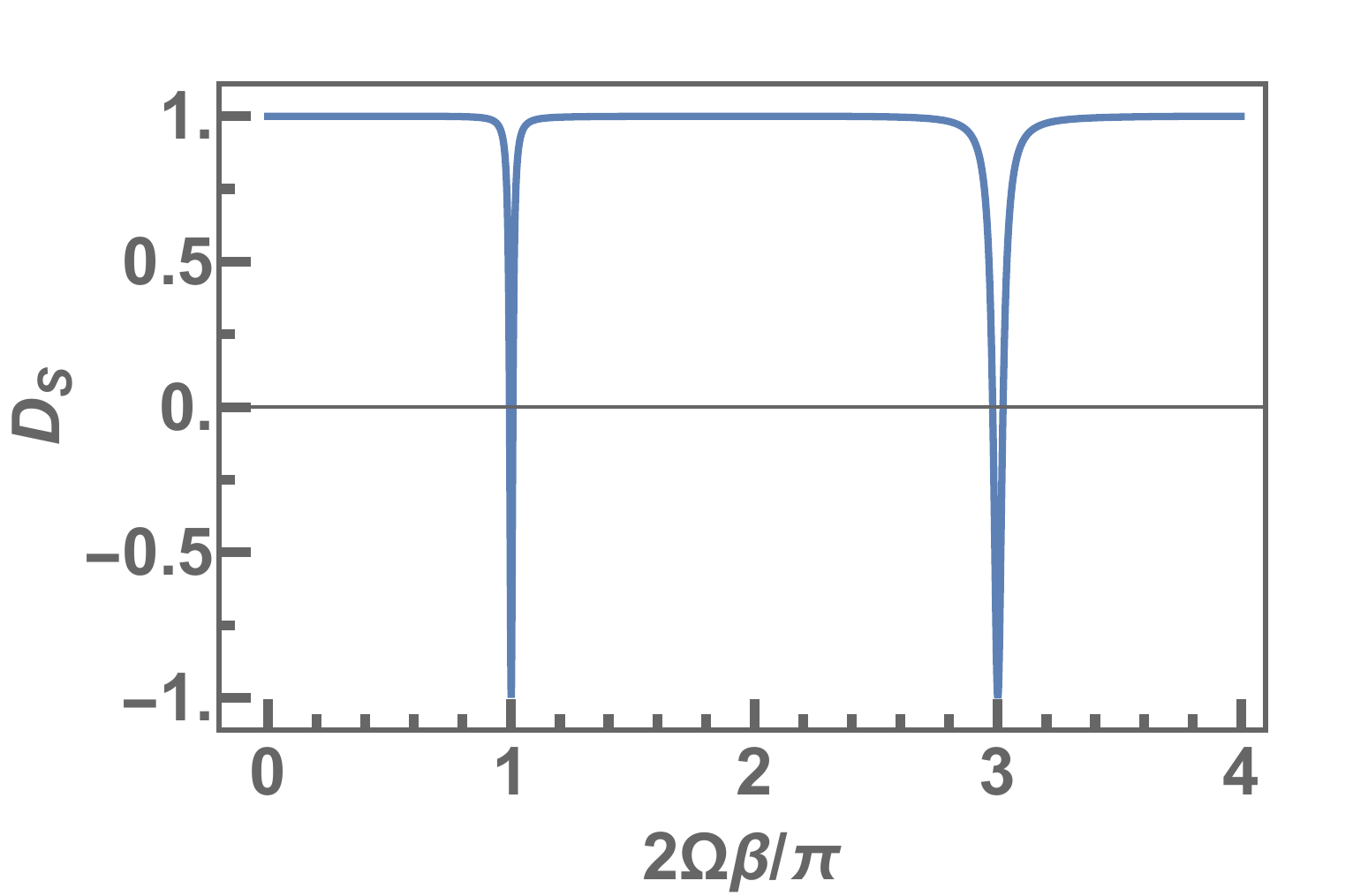}
        \caption{$2\Omega\beta/\pi\in[0,4]$}
        \label{fig:fig4a}
    \end{subfigure}
  \hfill
    \begin{subfigure}[b]{0.3\textwidth}
        \centering
        \includegraphics[width=\linewidth]{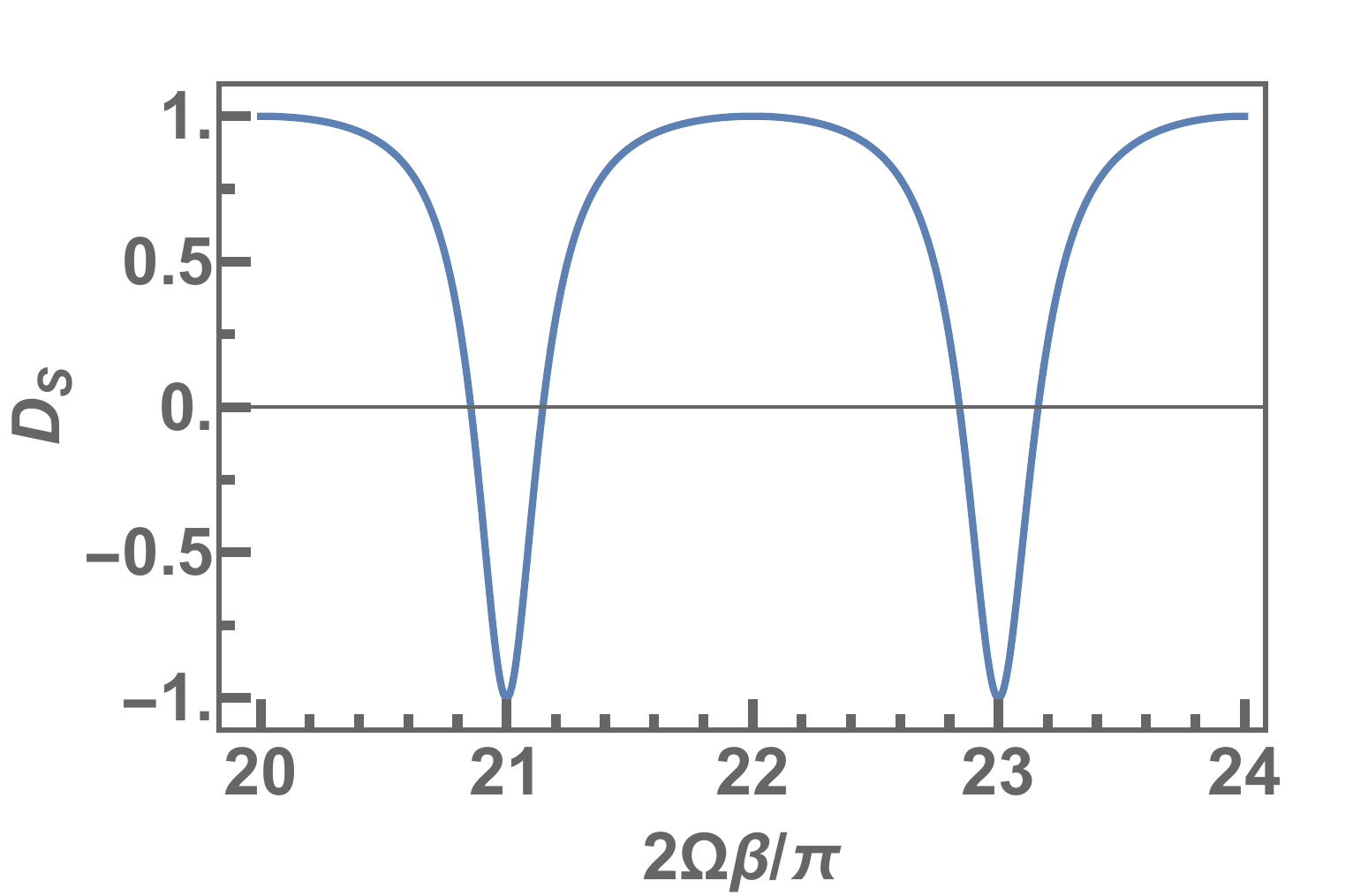}
        \caption{$2\Omega\beta/\pi\in[20,24]$}
        \label{fig:fig4b}
    \end{subfigure}
      \hfill
    \begin{subfigure}[b]{0.3\textwidth}
        \centering
        \includegraphics[width=\linewidth]{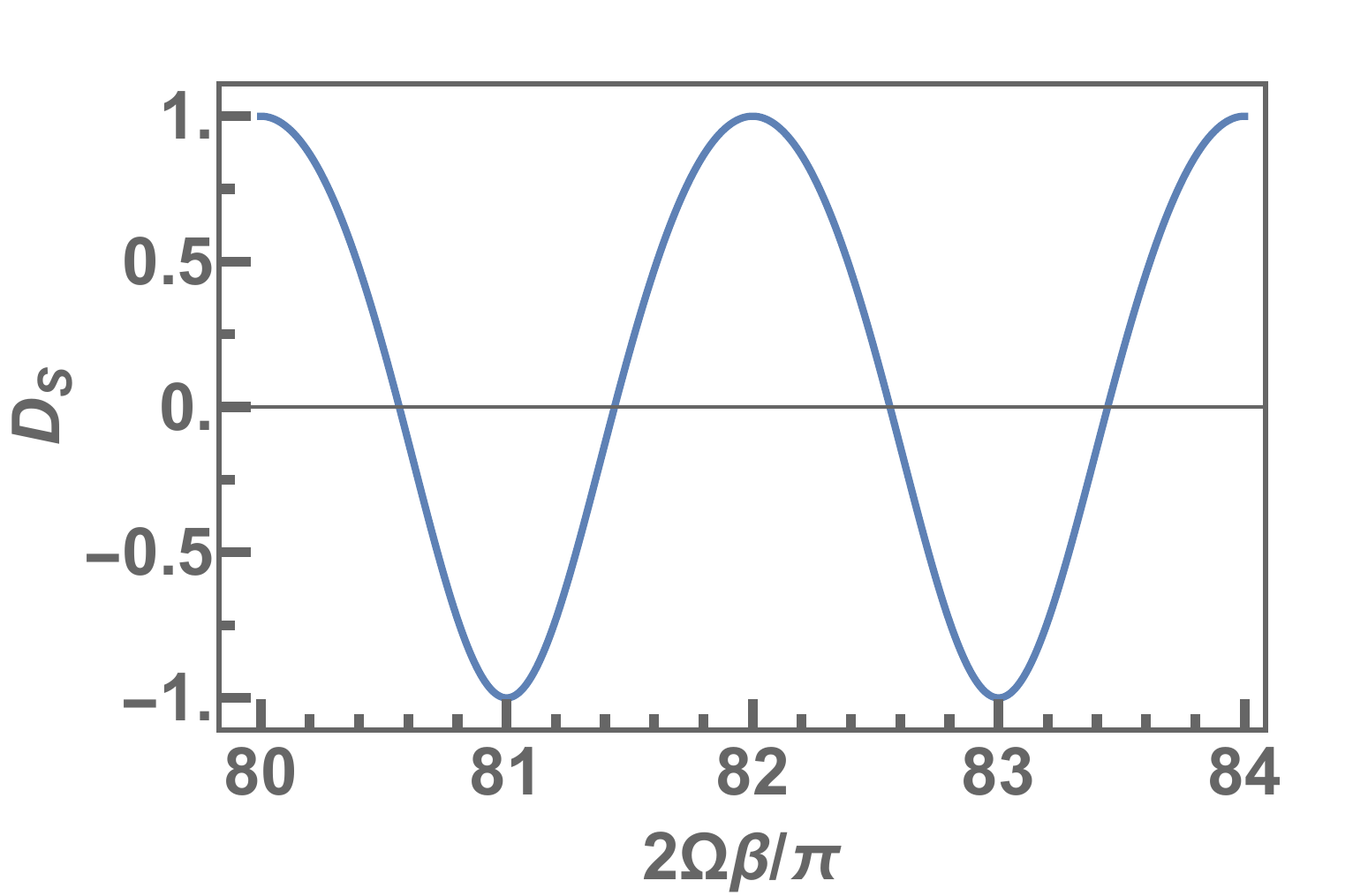}
        \caption{$2\Omega\beta/\pi\in[80,84]$}
        \label{fig:fig4c}
    \end{subfigure}

    \caption{Dependence of the symmetry degree $D_S$ on the time-delay parameter $\beta$
      computed for the biphoton state with the cosine-modulated JSA in the three different regions. 
      The parameters are:
      $\sigma_1 = \sigma_2 = 2\pi\times6$~THz,
      $2\Omega = 2\pi\times 844.5$~THz, $\sigma_p=0.01\sigma_1$, $\Delta\tau =0$.}
\label{fig:fig4} 
\end{figure*}

These effects are illustrated in
Fig.~\ref{fig:fig4}, which
shows the curves
representing the dependence of $D_S$ on $\beta$
in different ranges of the time-delay parameter.
The results
plotted in Fig.~\ref{fig:fig4}
are computed at
$\sigma_p = 0.01\sigma_1$ and $\Delta\tau = 0$.
The values of the other parameters $2\Omega=2\pi\times 844.5$~THz (corresponding wavelength is $355$~nm) and
$\sigma_1=\sigma_2=2\pi \times6$~THz (corresponding wavelength width is $10$~nm) are taken
from Ref.~\cite{Cromb2023}.

At these parameters,
the ratio $[\xi\pi/(2\Omega)]^2=(\beta_0\xi)^2$
can be estimated at about $2.5\times10^{-4}$.
So, at $2n+1<10$,
the approximation $\beta^2\xi^2\ll1$ remains applicable,
and an estimate of
the half width at half minimum (HWHM) of the resonant dip
at $\beta=\beta_0=\pi /(2\Omega)$
can be derived using expression~\eqref{eq:eq4.1.5}.

For the dimensionless half-width parameter
$\varepsilon$ related to
the value of the half minimum time-delay parameter
$\beta_H=\pi(1+\varepsilon)/(2\Omega)$, where
$D_S(\beta_H) =0$,
we have
$\varepsilon\approx\arccos{(1-\beta_0^2\xi^2)}/\pi\approx0.007$ producing the time-shift $\delta\beta=\varepsilon\pi/(2\Omega)\approx4$ as.
The corresponding
variation in the path-length difference of the MZI arms
is $\lambda\varepsilon/2$,
where $\lambda=2\pi c/\Omega$ is the wavelength.
In our case, it is about $2.5$~nm.
%(\textit{$\lambda/2\approx 1.5\times 10^{-7}$~m?})
This value indicates the high sensitivity of
the antibunching regime to fine adjustments of the optical path
illustrated in Fig.~\ref{fig:fig4a}.

By contrast,
measurements of the standard HOM interferometry
deal with the dependence of the coincidence probability (the symmetry degree),
$P_{2c}$ ($D_S$),
on the phase delay governing parameter $\Delta \tau$, which does not affect the entanglement degree.
In this case,
equation~\eqref{eq:eq2.2.11} can be used
to estimate the HWHM of the HOM dip for the SPDC biphotons with $\beta=0$.
So, we have the relation
$\Delta\tau_H=\sqrt{\ln{2}} \sqrt{\sigma_1^2+\sigma_2^2}/(\sigma_1\sigma_2)$
giving the estimate: $\Delta\tau_H\approx 30$~fs,
which is equivalent to the optical path
$c \Delta\tau_H$ of about
$9~\mu$m.

Despite noticeable sensing advantages offered by the modulated biphoton states,
  their metrological capabilities are constrained by the need to perform the post-selection
  procedure. Specifically,  the probability of projecting onto the
  cosine-modulated state near the resonant points $\beta \approx \beta_n$
can be estimated using formula~\eqref{eq:P_cond} that, under the condition
$\beta^2 \alpha^2 \ll 1$, assumes the simplified form:
$P_{\ind{cos}} \approx \beta^2 \alpha^2/2$.
  This value is typically quite small. For instance, at the first resonance shown in Fig.~\ref{fig:fig4a}
  ($\beta = \beta_0$), the probability is $P_{\ind{cos}} \approx 10^{-4}$ (HWHM is about $4$~as),
  implying that on average only one out of ten thousand experimental runs successfully yields
  the desired state. This probability, however, increases with the resonance number $n$.
  At $\beta=\beta_1$ and $n=1$,
  the probability estimate is
  $P_{\ind{cos}}\approx 10^{-3}$  (HWHM is about $13$~as).

  Note that the above estimates are valid for the spectral parameters listed in the
  caption of Fig.~\ref{fig:fig4} and another set of the parameters
  $\{\Omega,\sigma_1,\sigma_2,\sigma_p\}$
  will produce different results.
  In particular, an increase in the pump radiation central
  frequency, $2\Omega$, will reduce both HWHM and the post-selection probability because
  $\beta_n\propto\Omega^{-1}$ and $P_{\ind{cos}}\propto\beta$.

  So, given the experimental setup, it
  is essential to find an optimal resonance point $\beta_{n}$ that combines a reasonably small HWHM with a
  relatively high post-selection probability $P_{\ind{cos}}$.
  In our case, the point $\beta=\beta_1$ might serve the purpose.
The standard SPDC setups using common nonlinear crystals can produce biphoton
  pairs at a rate on the order of $10^4$ pairs per second~\cite{Lyons2018} that would yield
  several tens of cosine-modulated biphotons per second at $\beta=\beta_1$. More advanced biphoton
  sources, for example, reported in~Refs. \cite{Li2025, Niu2023, Appas2022,Autebert2016}, can
  increase this generation rate by several orders of magnitude, delivering up to $10^4$
  cosine-modulated biphotons per second. So, the challenge of low post-selection
  efficiency associated with narrow resonances does not pose an insurmountable barrier for metrology
  with the modulated biphoton states.

The low probability of success of the post-selection procedure
  applied to the state~\eqref{eq:eq4.1.2} may also have a detrimental effect on the precision of the
  modulation parameter estimation. Specifically,
according to the estimation theory~\cite{Helstrom:bk:1976}, if the cosine-modulated state is
  post-selected, the information carried by the sine-modulated state is lost leading to a reduction in the
  precision of the estimation of $\beta$.
  Roughly, given the total number of SPDC biphoton pairs $\mathcal{N}$, only
  the small fraction, $P_{\ind{cos}}\mathcal{N}$, will provide useful information for estimation.
  Below we perform a more accurate analysis and show that, in properly designed experiments,
  the estimation errors can be significantly reduced
even if the post-selection probability is low.
  % due to contributions of
  % all $\mathcal{N}$ pairs.
  
To this end, we consider the experimental scheme depicted in
  Fig.~\ref{fig:fig3}. In order to
  evaluate the precision of estimation of $\beta$
  we shall follow the standard approach previously used in
  the context of the HOM
  interferometry~\cite{Fabre2021,Jordan2022,Descamps2023,Meskine2024,Lyons2018}.
  So, the starting point of this approach is
  the  classical Cram\'{e}r–Rao inequility for the variance of an unbiased
  estimator, $\hat{\beta}$,
  of the modulation parameter~\cite{Helstrom:bk:1976,Nielsen2013}
\begin{align}
  &
    \Delta^2\hat{\beta} \ge \frac{1}{\mathcal{N}\mathcal{F}},
\label{eq:CRB}
\end{align}
where
$\Delta^2\hat{\beta}$ is the estimator variance;
$\mathcal{N}$ is the number of the experimental trials or the generated SPDC pairs,
and $\mathcal{F}$ is the classical Fisher information.
The right-hand side of Eq.~\eqref{eq:CRB} gives the ultimate
achievable limit on the precision of estimation known as the
Cram\'{e}r–Rao bound.
This limit is determined by the classical Fisher information (CFI)
that will be our primary concern in the remaining part of this subsection.
  
  First, we compute the CFI assuming that both the detectors $D_1$ and
  $D_2$ are photon number resolving.
  Since our task is to demonstrate the metrological capabilities of the modulated
  biphotons in principle, imperfections of the detector
  and decoherence effects will also be neglected.

Under these assumptions, the setup yields three outcomes per
  experimental trial: (a)~the detectors $D_1$ and $D_2$ simultaneously detect one photon with
  the probability $P_{2}=P_{\ind{cos}}(1-D_S)/2$;
  (b)~one of the detectors (either $D_1$ or $D_2$) detects two photons
  with the probability $P_{1}=P_{\ind{cos}}(1+D_S)/2$;
  and (c)~one of the detectors detects one photon
  with the probability $P_{\ind{sin}}=1-P_{\ind{cos}}$.
  Given the probabilities,  the CFI
\begin{align}
  &
    \mathcal{F}=\frac{1}{P_{1}}\left(\frac{\partial P_1}{\partial
    \beta}\right)^2+
    \frac{1}{P_{2}}\left(\frac{\partial P_2}{\partial \beta}\right)^2+
    \frac{1}{P_{\ind{sin}}}\left(\frac{\partial P_\ind{sin}}{\partial \beta}\right)^2
\label{eq:FI_3_events}
\end{align}
can be computed in the explicit form.
At $\sigma_p=0$ and $~\Delta\tau=0$,
the result reads
\begin{align}
    &\mathcal{F}=\frac{4e^{-2\beta^2\xi^2}[\beta\xi^2\cos{(2\beta\Omega)}+\Omega\sin{(2\beta\Omega)}]^2}{1-e^{-2\beta^2\xi^2}\cos^2{(2\beta\Omega)}}\notag\\
    &+\frac{2[\Omega(1-e^{-2\beta^2\xi^2})+\beta\xi^2e^{-\beta^2\xi^2}\sin{(2\beta\Omega)}]^2}{(1-e^{-2\beta^2\xi^2})[1+e^{-\beta^2\xi^2}\cos{(2\beta\Omega)}]}.
    \label{eq:FI_ideal_no_ps}
\end{align}
At the resonance points $\beta=\beta_n$,
where $\beta_n^2\xi^2\ll1$,
the CFI~\eqref{eq:FI_ideal_no_ps} can be estimed at about
$\mathcal{F}\approx 4\Omega^2+2\xi^2$.
This value is higher than the
maximal CFI for a non-modulated degenerate biphoton state in a HOM
interferometer that can be estimated at $2\xi^2$~\cite{Lyons2018}
by the factor $(\Omega/\xi)^2$.
For the parameters listed in the
caption of Fig.~\ref{fig:fig4}, the
ratio $(\Omega/\xi)^2$ is about $10^4$. Thus, in contrast to the expected
precision loss, the modulated biphotons can lead to a significant
enhancement in precision.

  Note that
  the same metrological performance can be achieved
  by placing an additional
  detector $D_3$ at the output port $3$ of BS$_2$ (see Fig.~\ref{fig:fig3})
  rather than using photon number resolving detectors.
  In this case,
% If $D_3$ is used while $D_1$ and $D_2$ cannot resolve
% photon-number,
  the event probabilities remain unchanged
  and
  the outcomes are: (a)~$D_1$ and $D_2$ click simultaneously;
  (b)~$D_1$ (or $D_2$) clicks; and
  (c)~$D_3$ and $D_1$ (or $D_2$) click.

In the case, where additional detector is not used and the detectors
$D_1$ and $D_2$ cannot
resolve photon numbers,
measurements cannot distinguish
between the above listed events (b) and (c)
and
the number of the experimental
outcomes is reduced to two:
(a)~both detectors click with the probability $P_{2}$;
(b)~ either $D_1$ or $D_2$ clicks with the probability
$\tilde{P}_{1}=P_{1}+P_{\ind{sin}}$.
Now the CFI assumes the form
\begin{align}
  &
    \tilde{\mathcal{F}}=\frac{1}{\tilde{P}_{1}}\left(\frac{\partial
    \tilde{P}_1}{\partial
    \beta}\right)^2+\frac{1}{P_{2}}\left(\frac{\partial P_2}{\partial
    \beta}\right)^2
\label{eq:FI_postselect}
\end{align}
and, in the ideal scenario with $\sigma_p=0$ and $\Delta\tau=0$,
leads to the following expression:
\begin{align}
  &
    \tilde{\mathcal{F}}=
    \frac{2[\Omega\cos(\beta\Omega)(1-e^{-\beta^2\xi^2})+\beta\xi^2\sin(\beta\Omega)e^{-\beta^2\xi^2}]^2}{1-e^{-\beta^2\xi^2}}
    \notag
  \\
  &
    +\frac{[\Omega\sin(2\beta\Omega)(1-e^{-\beta^2\xi^2})-2\beta\xi^2(\cos^2(\beta\Omega)-e^{-\beta^2\xi^2})]^2}{%
    2[\cos^2(\beta\Omega)(1-e^{-\beta^2\xi^2})+e^{-\beta^2\xi^2}]}.
    \label{eq:FI_ideal_ps}
\end{align}
At the resonance points $\beta=\beta_n$ with $\beta_n^2\xi^2\ll1$,
formula~\eqref{eq:FI_ideal_ps} gives the estimate
$\tilde{\mathcal{F}}\approx2\xi^2$
that coincides with
the value of the maximal CFI provided by non-modulate
degenerate biphotons in HOM interferometer~\cite{Lyons2018}.

%%%%%%%%%%%%%%%%%%%%%%%%%
\subsection{Entanglement properties}
\label{subsec:entangl}
%%%%%%%%%%%%%%%%%%%%%%%%%

The Schmidt decomposition of the cosine-modulated state can be
performed based on
the previously discussed Mehler formula~\eqref{eq:eq3.1.2} that leads to the decomposition:
\begin{align}
  &
    \left|\Psi \right\rangle =
    P_{cos}^{-1/2}\sum_{n=0}^{\infty} \sqrt{\lambda_n}|\tilde{\phi}_n \rangle_1|\vartheta_n \rangle_2,
    \label{eq:eq4.1.6}
  \\
  &
    |\tilde{\phi}_n \rangle_1 = \frac{1}{\sqrt{s_1}}
    \int_{-\infty}^{\infty}d\omega_1\psi_n(\tilde{\omega}_1/s_1)e^{i\omega_1\tau_1}
    \notag
  \\
  &
    \times
    \cos{(\beta\omega_1)}\hat{a}_1^{\dagger}(\omega_1)|0\rangle_1.
\label{eq:eq4.1.7}
\end{align}
Note
that Eq.~\eqref{eq:eq4.1.6} cannot be treated as the Schmidt
decomposition since the vectors~\eqref{eq:eq4.1.7} are not
orthogonal. However, it can be used for both numerical and analytical
derivations of the Schmidt number. For these purposes, it is
convenient to exploit the reduced density matrix, given by
\begin{align}
&\hat{\rho}_1=P_{cos}^{-1}\sum_{n=0}^{\infty} \lambda_n|\tilde{\phi}_n \rangle_1\langle\tilde{\phi}_n|.
\label{eq:eq4.1.8}
\end{align}
The elements of the reduced density matrix $\hat{\rho}_1^{pq}$ can be found
in the Fock basis defined by the states
$\left\{|\phi_n \rangle\right\}$ provided in \eqref{eq:eq3.1.3}. The
relevant scalar product $\langle\phi_p|\tilde{\phi}_n \rangle$ is then
given by
\begin{align}
  \langle\phi_p|\tilde{\phi}_n \rangle =&\frac{e^{i\beta\Omega}+(-1)^{p+n}e^{-i\beta\Omega}}{2} \notag\\
&\times                                          G_{pn}\left(\frac{\beta^2 s_1^2}{2}\right)e^{-\beta^2 s_1^2/4},
\label{eq:eq4.1.9}
\end{align}
where $G_{pn}(x)=\sqrt{r!/s!}(i\sqrt{x})^{s-r}L_r^{(s-r)}(x)$,
$L_r^{(s-r)}(x)$ is the associated
Laguerre polynomial, $s=\max(p,n)$, $r=\min(p,n)$.
Derivation of the scalar
product~\eqref{eq:eq4.1.9} is detailed in Appendix~\ref{apendA}.
Thus,
the reduced density matrix elements, $\hat{\rho}_1^{pq}$, can be calculated via the relation
\begin{align}
  \label{eq:eq4.1.10}
%  \label{eq:query}
  &
    \hat{\rho}_1^{pq}=P_{cos}^{-1}\sum_{n=0}^{\infty} \lambda_n
    \langle\phi_p|\tilde{\phi}_n \rangle
    \langle\phi_q|\tilde{\phi}_n \rangle^*.
\end{align}

The eigenvalues of the reduced density matrix $\hat{\rho}_1$ correspond to
the Schmidt coefficients $\tilde{\lambda}_n$.
(In Sec.~\ref{sec:sec3},
the eigenvalues associated with the standard
biphoton state are denoted by 
$\lambda_n$.)
Thus,
the Schmidt number of the cosine-modulated state
can be evaluated using numerical diagonalization of the reduced
density matrix.

On the other hand, an analytical expression for the
Schmidt number would facilitate a deeper and more efficient
analysis of the relationship between entanglement and
antibunching. Unfortunately, direct extraction of the Schmidt modes
for the cosine-modulated biphoton state poses considerable analytical
challenges.
In order to circumvent this difficulty, we adopt the perturbation
theory approach combined with some
specific heuristic procedure described
in Appendix~\ref{apendB}.
This method gives the following analytical result for the Schmidt number
\begin{align}
  &
    \tilde{K}\approx K_0
    \frac{\big(1+e^{-\beta^2\alpha^2}\cos{(2\beta\Omega)}\big)^2}{1+2R+R^2I_0\left(4\eta^2z^2\right)},
    \label{eq:eq4.1.11}
\end{align}
where $K_0$ is given by~\eqref{eq:eq3.1.4},
$R = \cos{(2\beta\Omega)}e^{-\eta^2(1+z^4)}$,
$\eta^2 = \beta^2s_1^2/(1-z^4)$, and $I_0(x)$ is the modified Bessel
function.
Details on
the derivation of
the approximate Schmidt number~\eqref{eq:eq4.1.11} and its accuracy are relegated
to Appendix~\ref{apendB}.

Note that in the limiting case, where $\sigma_p\rightarrow0$ and $\beta^2\xi^2\ll1$,
the expression for the Schmidt number assumes the simplified form:
\begin{align}
  &
    \tilde{K}\approx K_0^{\sigma_p\rightarrow0}\bigl(1+\cos{(2\beta\Omega)}-\beta^2\xi^2\cos{(2\beta\Omega)}\bigr)^2
    \notag
  \\
  &
    \times
    \Bigl[(1+\cos{(2\beta\Omega)}-\beta^2\xi^2\cos{(2\beta\Omega)}/2\big)^2
    \notag
  \\
  &
    +\cos^2{(2\beta\Omega)}\beta^4\xi^4/4\Bigr]^{-1},
    \label{eq:eq4.1.12}
\end{align}
where we have used
the approximation for the modified Bessel function $I_0(x)\approx 1+x^2/4$ valid at small $x$.
In this case, the Schmidt number $K$ remains close to its baseline value
$K_0^{\sigma_p\rightarrow0}$ for all $\beta$,
except the close vicinity of the
resonance points $\beta_n=\pi(2n+1)/(2\Omega)$, where it doubles to
$K=2K_0^{\sigma_p\rightarrow0}$.
Thus, similar to
the symmetry degree, $D_S$,
the $\beta$-dependence of the Schmidt number,
$K$, is expected to show
sharp and narrow resonant peaks.
This behavior highlights a striking
correlation between antibunching and the degree of entanglement.

Formulas~\eqref{eq:eq4.1.4},~\eqref{eq:eq4.1.10} and~\eqref{eq:eq4.1.11}
enable us to fully uncover the interrelation
between the two-photon coincidence probability and the degree of
entanglement in the modulated biphoton state. Since the symmetry
degree of the modulated state can be controlled via the parameter
$\beta$, it is natural to compare
the dependencies of $D_S$ and $K$ on this time-delay parameter.

\begin{figure*}
\centering
    \begin{subfigure}[b]{0.3\textwidth}
        \centering
        \includegraphics[width=\linewidth]{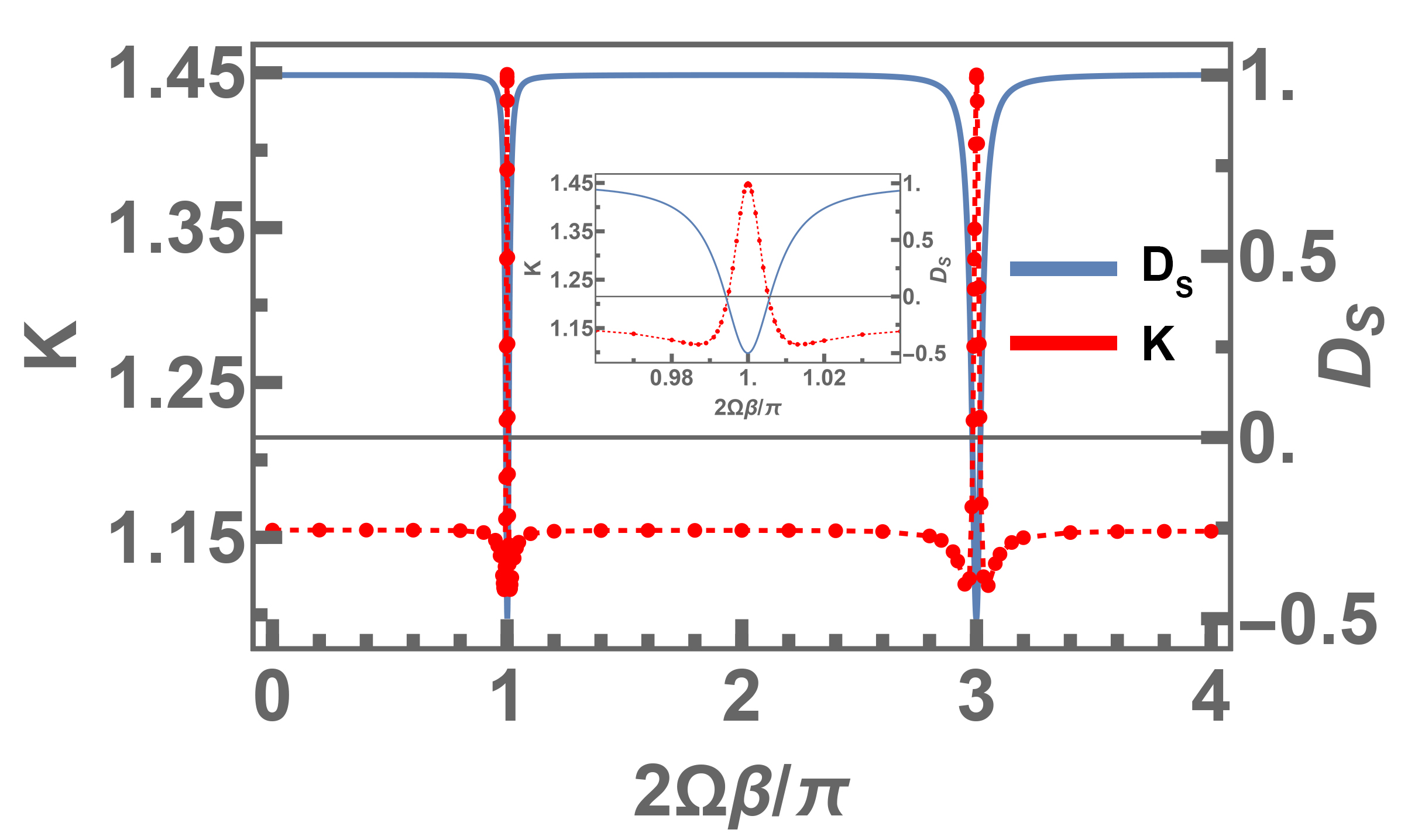}
        \caption{$\sigma_p=\sigma_1,~2\Omega\beta/\pi\in[0,4]$}
        \label{fig:fig5a}
    \end{subfigure}
   \hfill
    \begin{subfigure}[b]{0.305\textwidth}
        \centering
\includegraphics[width=\linewidth]{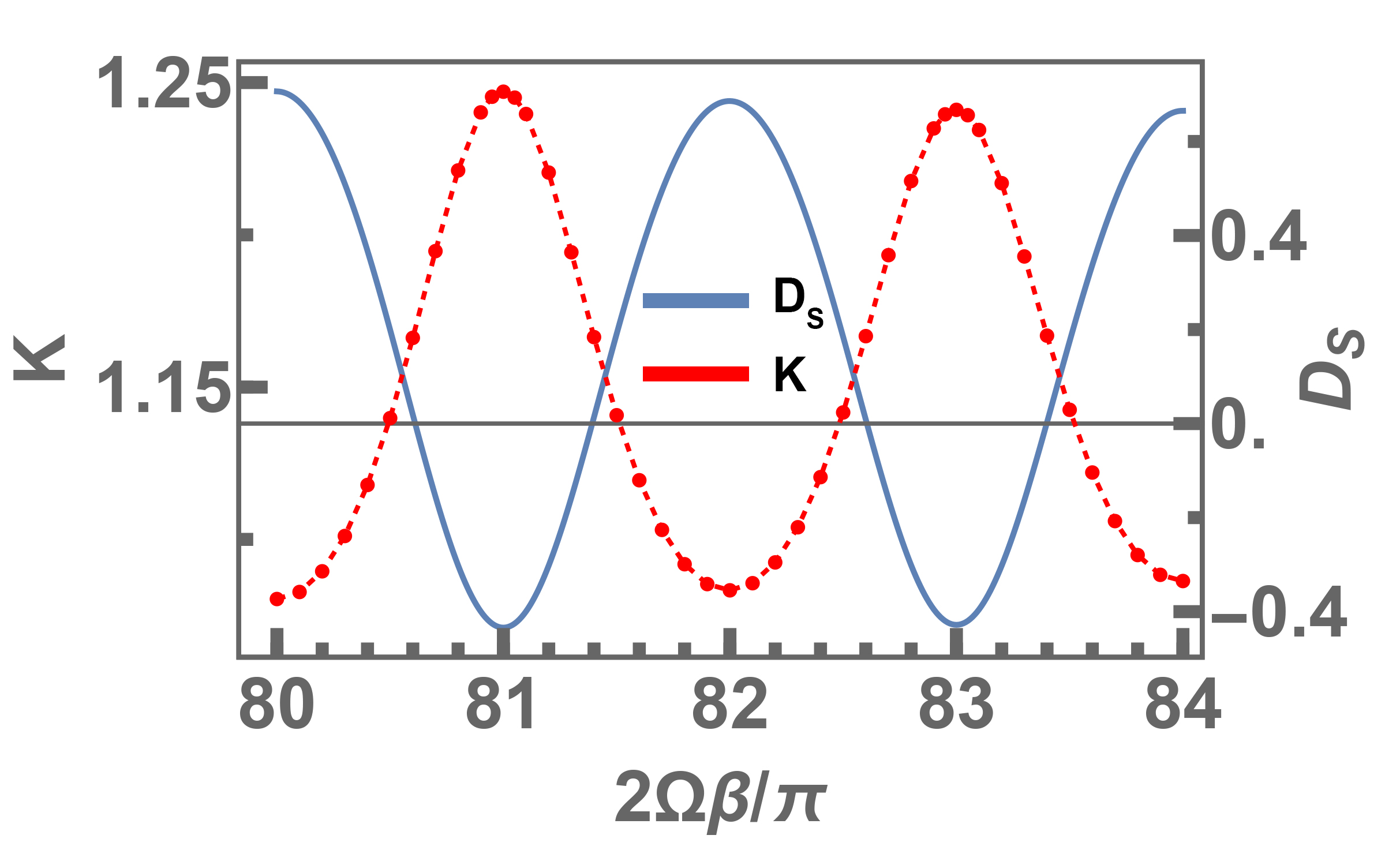}
        \caption{$\sigma_p=\sigma_1,~2\Omega\beta/\pi\in[80,84]$}
        \label{fig:fig5b}
    \end{subfigure}
    \hfill
    \begin{subfigure}[b]{0.295\textwidth}
        \centering
        \includegraphics[width=\linewidth]{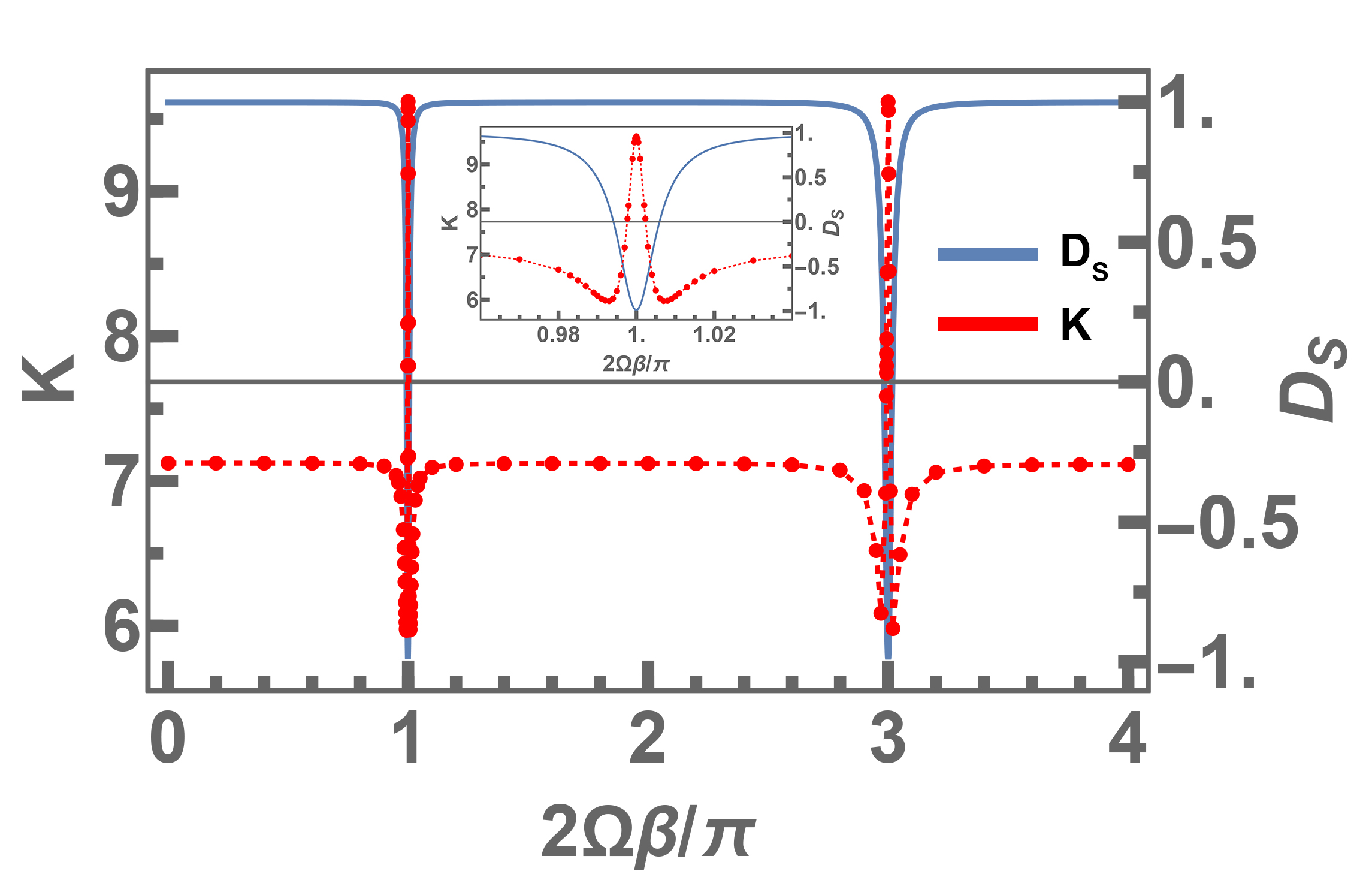}
        \caption{$\sigma_p=0.1\sigma_1,~2\Omega\beta/\pi\in[0,4]$}
        \label{fig:fig5c}
    \end{subfigure}
      \vspace{0.5em} 
      
    \begin{subfigure}[b]{0.3\textwidth}
        \centering
        \includegraphics[width=\linewidth]{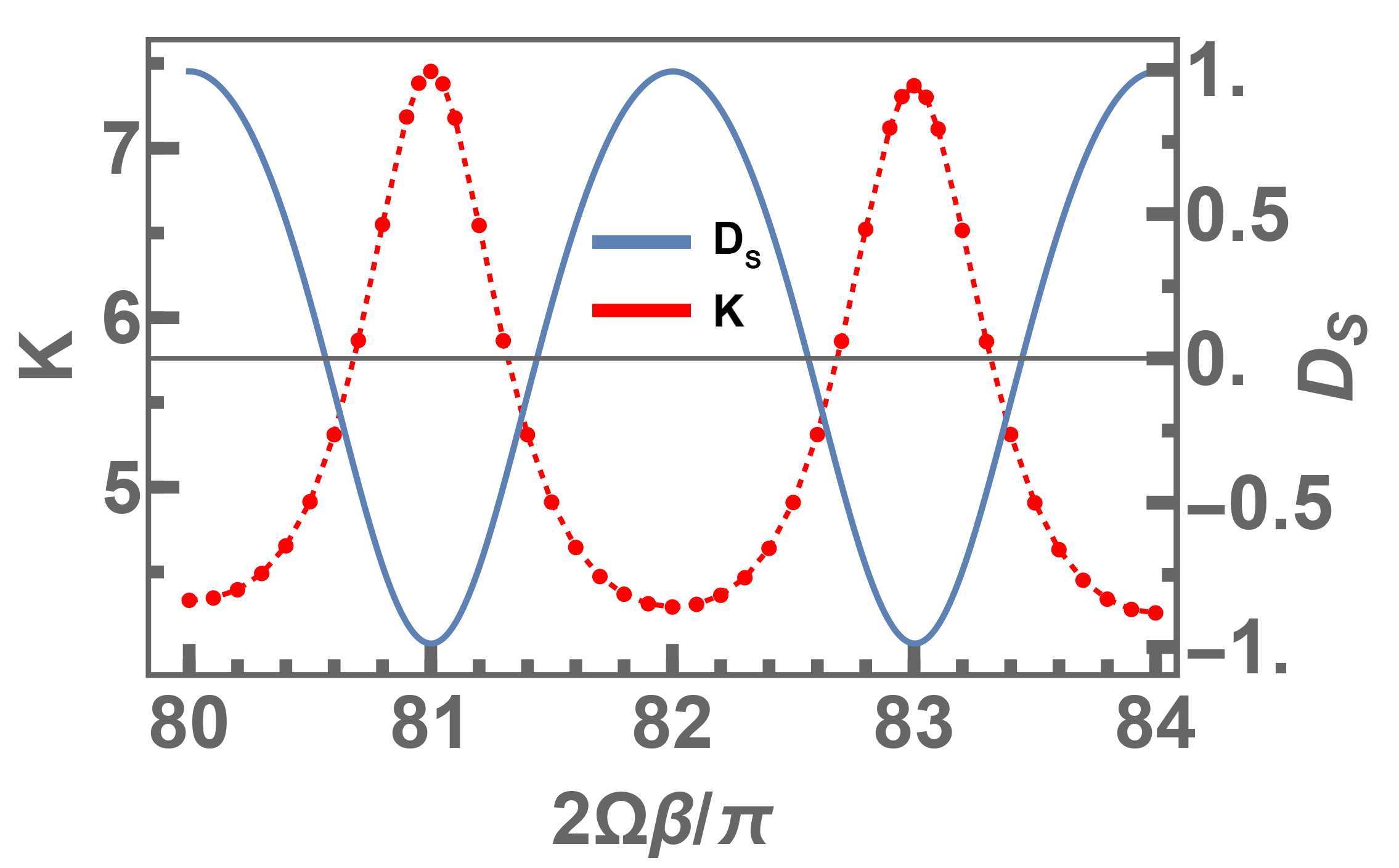}
        \caption{$\sigma_p=0.1\sigma_1,~2\Omega\beta/\pi\in[80,84]$}
        \label{fig:fig5d}
    \end{subfigure}
     \hfill
    \begin{subfigure}[b]{0.305\textwidth}
        \centering
        \includegraphics[width=\linewidth]{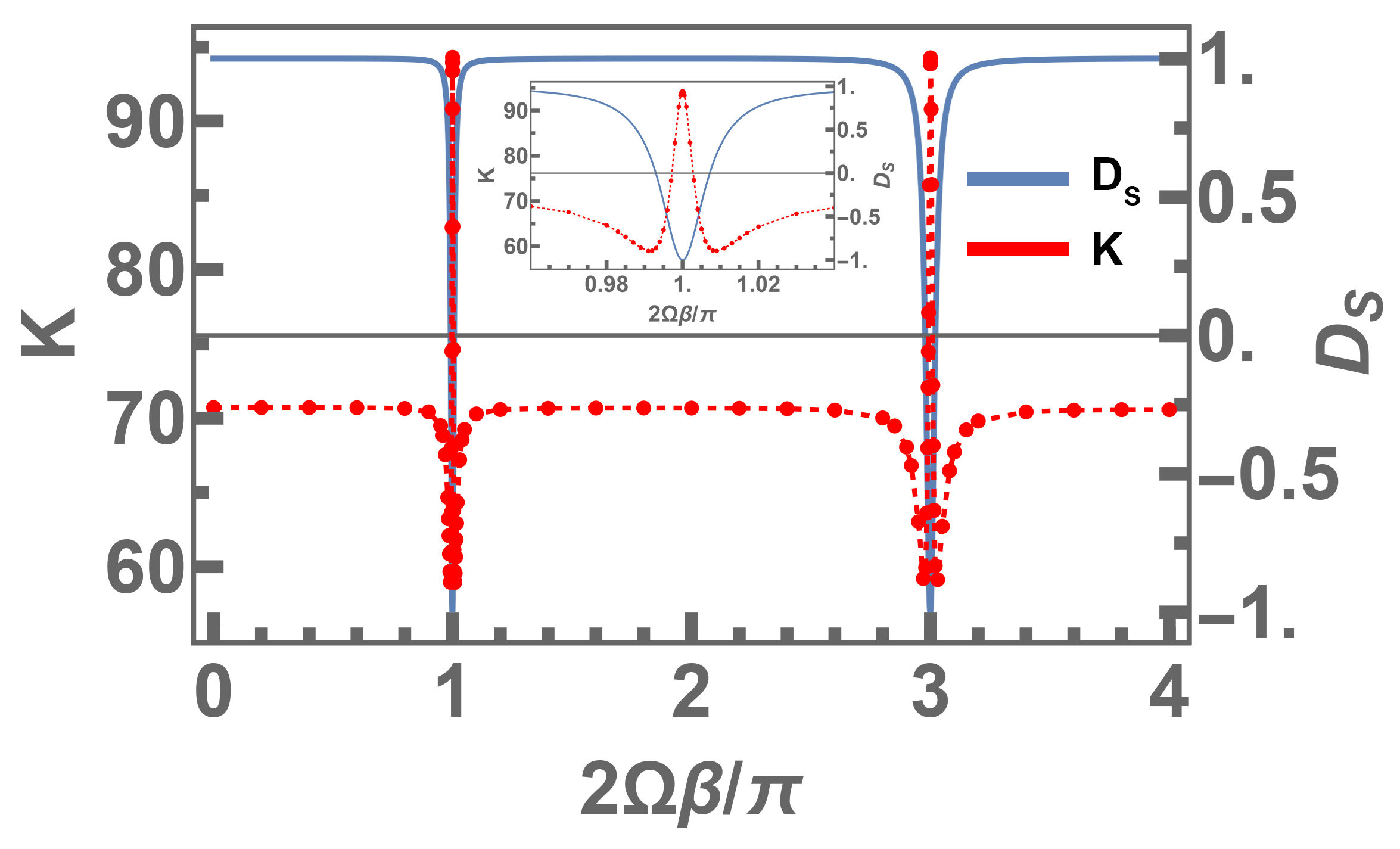}
        \caption{$\sigma_p=0.01\sigma_1,~2\Omega\beta/\pi\in[0,4]$}
        \label{fig:fig5e}
    \end{subfigure}
     \hfill
    \begin{subfigure}[b]{0.3\textwidth}
        \centering
        \includegraphics[width=\linewidth]{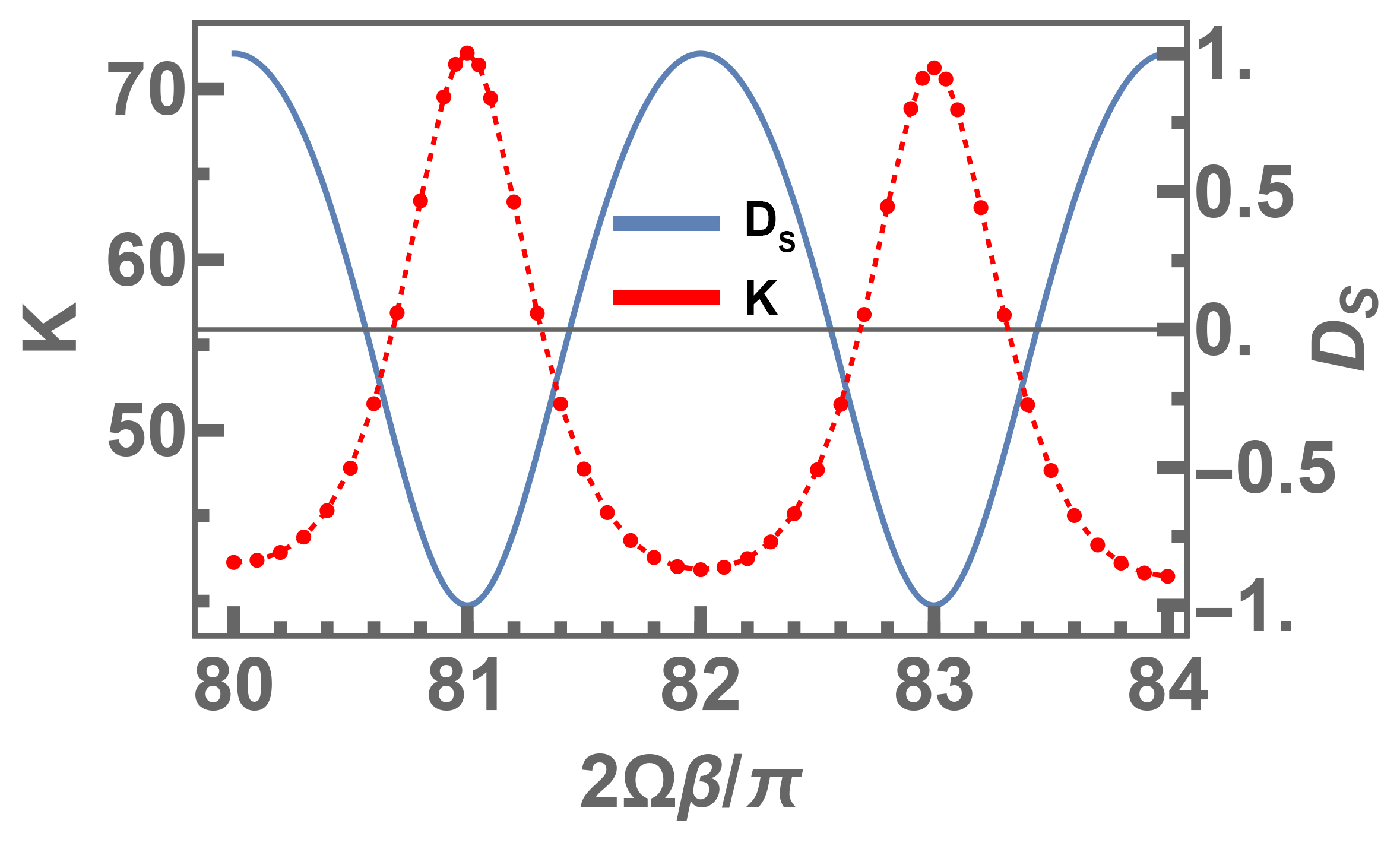}
        \caption{$\sigma_p=0.01\sigma_1,~2\Omega\beta/\pi\in[80,84]$}
        \label{fig:fig5f}
    \end{subfigure}

    \caption{The Schmidt number, $K$,
      and the symmetry degree, $D_S$, as a function of the time-delay parameter $\beta$
      for the biphoton state with the cosine-modulated JSA.
      The parameters are listed in the caption of
      Fig.~\ref{fig:fig4}.
      The insets display a zoomed-in view of the first resonant peak in the
      Schmidt number and the symmetry degree profiles.}
\label{fig:fig5} 
\end{figure*}

Figure~\ref{fig:fig5} presents these dependencies computed
using Eqs.~\eqref{eq:eq4.1.4} and~\eqref{eq:eq4.1.10}, for
$\sigma_p = \{\sigma_1, 0.1\sigma_1, 0.01\sigma_1\}$
at $\sigma_1 = \sigma_2$ and $\Delta\tau = 0$.
A comparison between
the curves shown in Fig.~\ref{fig:fig5}
suggests a strong
correlation between $D_S(\beta)$ and $K(\beta)$.
It is seen that the loci of the maxima (minima) of the
Schmidt number
coincide with those of the minima (maxima)
of the symmetry degree
(the maxima (minima) of the two-photon coincidence probability, $P_{2c}$).
As
predicted, for small values of $\beta$, sharp narrow resonances for
the symmetry degree and the Schmidt number are
observed. Hence,
the $\beta$-dependencies of both $D_S$ and $K$
appear to be highly sensitive to variations in the time-delay
parameter.
These results highlight how the modulated biphoton states can be
employed for precision measurements. Figures~\ref{fig:fig5} and~\ref{fig:fig4} also
  demonstrate another non-trivial fact that, for the
  cosine-modulated biphotons,
  the parity and perfect (anti)bunching conditions (Eqs.~\eqref{eq:eq2.2.9} and~\eqref{eq:perfect_antibunch})
determine the loci of extrema of these dependencies.

Despite the strong similarity between
the $\beta$-dependencies of $D_S$ and $K$,
the Schmidt number exhibits a distinctive feature not
present in the symmetry degree: wing-like dips near the entanglement
maxima.
In these regions, the degree of entanglement drops sharply
below its value at points corresponding to highly positive
symmetry. Notably, the depth of these dips increases with
the degree of entanglement that may have a detrimental effect on
the accuracy of
the Schmidt number determination from the measured values of $D_S$.

\begin{figure}
\centering 
\includegraphics[width=1\linewidth]{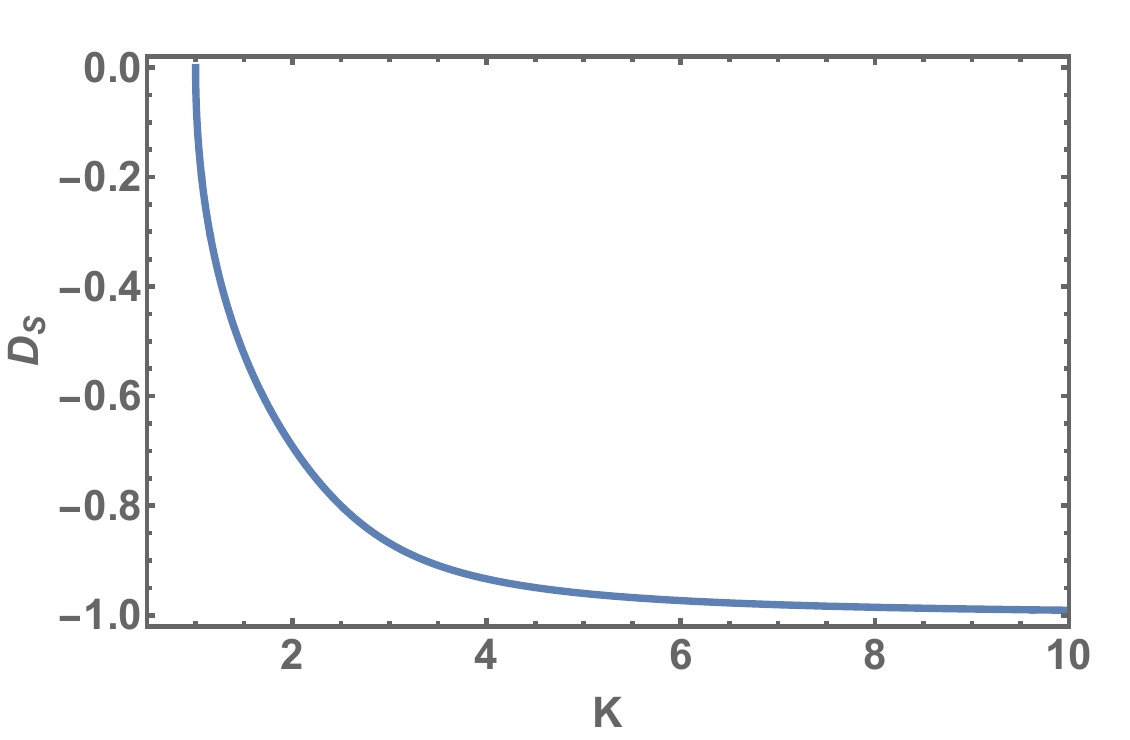}
\caption{The symmetry degree at zero-order resonance with $\beta=\beta_0=\pi/(2\Omega)$,
  $D_S(\beta_0)$, as a function of the Schmidt number, $K$,
  for the biphoton state with the cosine-modulated JSA.  
}
\label{fig:fig6}
\end{figure}

Note that
one of the key factors
limiting this accuracy is  the sensitivity of the symmetry degree to the Schmidt number.
In Fig.~\ref{fig:fig6},
the zero-resonance value of the symmetry degree,
$D_S(\beta_0)$,
is plotted against the Schmidt number, $K$.
It is clearly seen that $D_S(\beta_0)$
saturates, approaching
the perfect antibunching limit with $D_S=-1$ as 
the Schmidt number increases.
Referring to Fig.~\ref{fig:fig6},
in the region of highly entangled states with $K>8$,
owing to low sensitivity of $D_S$ to $K$,
the symmetry degree
will fail to provide sufficiently accurate quantitative information
about the entanglement degree of such states.

The results of this section suggest
  that, for the modulated biphotons,
  the behavior of the curves for the Schmidt number, $K$,
  and the symmetry degree, $D_S$, is somehow correlated:
  $K$ increases (decreases) when $D_S$ decreases (increases).
  Variations of the Schmidt number are induced by modulation
  arising from a non-unitary transformation (filtering process) performed within the
  Mach–Zehnder interferometer (see Sec.~\ref{subsec:generation}).
  The question of whether this type of modulation may result in similar correlation between $D_S$ and $K$ for
  differently shaped biphotons remains open and is the subject of future studies.

%%%%%%%%%%%%%%%%%%%%%
\section{Discussions and conclusion}
\label{sec:disc-concl}
%%%%%%%%%%%%%%%%%%%%%%%%

In this work, we have studied the spectral properties of biphotons that govern both the bunching and
antibunching regimes of the HOM two-photon interferometer, along with the quantum entanglement of
the SPDC biphoton 
states.

  Our analysis is concentrated on
  the specific family of experimentally realizable SPDC-generated biphoton states
determined by the JSA of the form given by Eq.~\eqref{eq:eq2.2.1}.
In order to quantify the antibunching effect,
we have expressed the two-photon coincidence probability~\eqref{eq:eq2.1.12},
$P_{2c}$, in terms of the symmetry degree parameter, $D_S$, (see
Eq.~\eqref{eq:eq2.1.13}) determined by the biphoton state joint spectral
amplitude (JSA). Specifically, negative values of this parameter indicate antibunching, whereas in the opposite
case of bunching, the symmetry degree is positive.  

Formulas~\eqref{eq:eq2.1.12} and~\eqref{eq:eq2.1.13} both suggest
that the type of the two-photon interference regime depends on
the balance between the symmetric and antisymmetric parts of the JSA.
In previous studies~\cite{Boucher2015,Fabre2020},
the biphoton states with
the antisymmetric JSA, such as the
so-called Schr\"odinger cat-like biphotons,
were engineered to demonstrate the antibunching effect.
For the physically feasible JSA taken in the alternative  form given by Eq.~\eqref{eq:eq2.2.1},
we have found that
the key factors dictating the sign of $D_S$ are related
to the parity properties of the \textcolor{magenta}{local} spectral function~\eqref{eq:phi_12}
whose even (odd) component (see Eq.~\eqref{eq:eq2.2.6})
determines the positive (negative) part of the symmetry degree~\eqref{eq:eq2.2.7}.
In particular,
the biphoton JSA
must satisfy the odd (even) parity condition given by Eq.~\eqref{eq:eq2.2.9}
to demonstrate the regime of antibunching (bunching) for any $\sigma_p<\infty$,
where the positive (negative) contribution to $D_S$ is suppressed:
$D_{S}^{(+)}=0$ ($D_{S}^{(-)}=0$).

Interestingly, the family of SPDC biphotons under consideration
  cannot produce the perfect antibunching with $D_S=-1$, until
  the pump variance approaches zero, $\sigma_p\to 0$.
This limitation arises because, at $\sigma_p>0$,
  the JSA~\eqref{eq:eq2.2.1} cannot be antisymmetric,
  $\psi(\omega_1,\omega_2)\not=-\psi(\omega_2,\omega_1)$. In the limit
  $\sigma_p\to 0$, the antisymmetry condition,
  $\psi(\omega_1,\omega_2)=-\psi(\omega_2,\omega_1)$, reduces to
  the form given by Eq.~\eqref{eq:perfect_antibunch}, thereby enabling the perfect antibunching.
  It is noteworthy that the latter condition is not equivalent to the odd-parity
  condition~\eqref{eq:eq2.2.9}.
  This reflects the fact that, in our model, the parity and symmetry
  properties of the JSA influence the symmetry degree in different ways. In particular, the parity
  conditions govern the sign of $D_S$,
  whereas the symmetry conditions ensure the
  extreme values of $D_S$. Each of these conditions can be satisfied independently.

Since the standard Gaussian SPDC-generated biphoton with the JSA~\eqref{eq:eq2.1.2}
cannot exhibit antibunching (see Eq.~\eqref{eq:eq2.2.11}),
we have introduced an
experimentally feasible family of modulated biphoton states that
can be tuned to meet the parity conditions.
These modulated states can be experimentally
realized by incorporating a Mach–Zehnder interferometer (MZI) or a Michelson interferometer into one of
the optical paths of the HOM interferometer (see Fig.~\ref{fig:fig3}) and
constitute a practical route to engineer spectral odd parity in
otherwise even parity SPDC outputs.
Mathematically, the modulation procedure leads to
the standard Gaussian SPDC JSA
modulated using harmonic functions 
(the modulating factor is either $\sin{(\beta\omega)}$ or
$\cos{(\beta\omega)}$) that depend on the time-delay parameter
$\beta=\Delta L/(2c)$.
The latter can be regarded as the governing parameter,
which is determined by the MZI path-length difference,
$\Delta L$,
and can be adjusted to switch between
the bunching and antibunching regimes (see Fig.~\ref{fig:fig4}).
Since antibunching is suppressed for separable states
(see Eq.~\eqref{eq:eq2.2.10}),
the transition between these regimes inevitably involves variation
in the degree of entanglement  of the modulated biphoton states.

In order to quantify the degree of entanglement of biphoton states,
we have applied the Schmidt decomposition technique.
For  the standard SPDC-generated Gaussian states~\eqref{eq:eq2.1.2},
similarly to the symmetry degree~\eqref{eq:eq2.2.11},
the Schmidt number, $K$, can be computed in the closed analytical form
given by Eq.~\eqref{eq:eq3.1.4}.
Though these states cannot show antibunching,
their analytical tractability renders them useful
for benchmarking more complex modulated states capable of exhibiting
negative $D_S$.

By extending our analysis to modulated biphoton states,
we have computed both the symmetry degree
(see Eq.~\eqref{eq:eq4.1.4})
and the Schmidt number as functions of
the time-delay parameter $\beta$.
The curves presented in Fig.~\ref{fig:fig5}
clearly demonstrate nontrivial and strong correlations between entanglement and
antibunching.

Specifically, in the region of small values of $\beta$, the $\beta$-dependencies of both $K$ and
$D_S$ exhibit sharp and narrow resonance peaks and dips corresponding to the maxima of the Schmidt
number and minima of the symmetry degree (maxima of the two-photon coincidence probability),
respectively. HWHM of the resonant dips translates into optical path variations on the nanometer
scale or time delays on the order of a few attoseconds. These values lie within the sensitivity
range of tactical-grade quantum optical gyroscope systems, underscoring the potential of such
quantum interference effects for precision sensing applications. For comparison, HWHM of the HOM
dip, resulting from the measurements of the two-photon coincidence probability as a function of
relative phase shift, $\Delta\tau$, corresponds to time delays on the order of tens of
femtoseconds. Thus, HOM interferometry with modulated biphotons in the antibunching regime appears
to be more sensitive than conventional HOM interferometry with standard SPDC biphotons in the
bunching regime.
  
However, as detailed in
    Sec.~\ref{subsec:generation}, the metrological capabilities of modulated biphotons engineered via MZI
    are limited by the low efficiency of the post-selection procedure.
    It turns out that the narrower the antibunching resonance, the smaller the success probability.
    For instance, at the first resonance ($\beta=\beta_0$), where
    HWHM is approximately $4$~as, the probability to obtain a cosine-modulated biphoton is
    $P_{\ind{\cos}} \approx 10^{-4}$. A more viable candidate for metrology is the third
    resonance ($\beta=\beta_1$) with a broader HWHM of about $13$~as
    and an approximately ten times enhanced value of the success probability
    $P_{\ind{\cos}} \approx 10^{-3}$.
    Since modern SPDC sources can deliver up to $10^7$ biphoton pairs per second,
    this probability is sufficient in practice.
    In addition,
    searching for a more efficient method to generate the modulated biphotons
may further improve the situation.

Another problem is that low probability of
   success of the post-selection
   procedure may result in a reduction in precision of modulation
   parameter estimation caused by information loss.
   It turned out that this problem is not insurmountable.
   In Sec.~\ref{subsec:symmdeg}, we have shown that,
   when the experiment is properly designed,
   the classical Fisher information for
   modulated biphotons is up to four orders of magnitude higher
   than that reported in~\cite{Lyons2018} for non-modulated degenerate SPDC biphotons.
   This implies 
   that the precision of measurements using modulated biphotons as a
   probe state determined by the Cram\'{e}r-Rao bound
   can also be up to four orders of magnitude higher than that
   achieved with standard SPDC biphotons. This highlights the
   remarkable sensing capabilities of modulated biphotons
   that will be analyzed in more detail in our subsequent publications.

 Apart from the sensitivity enhancement, antibunching resonances can
 be exploited for entanglement control -- a feature unattainable in
 conventional HOM dip measurements since the Schmidt number is
 insensitive to the relative phase delay $\Delta\tau$,
which is induced by local unitary
   transformations. By contrast, for the modulated biphoton states,
   variation of the entanglement degree with the time-delay parameter
   $\beta$ originates from a non-unitary filtering transformation
   performed by MZI.

    As we have already noted at the end of Sec.~\ref{subsec:entangl},
    the curves presented in Fig.~\ref{fig:fig5} reveal that,
    for the symmetry degree and the Schmidt number
    of the modulated biphotons,
    the loci of the extrema appear to be
    strongly correlated.
    An interesting and open
    question is whether such correlation can be found for differently shaped biphoton states.

As far as estimation of the entanglement degree is concerned,
our analysis has shown that the $D_S$-vs-$K$ dependence
quickly saturates (see Fig.~\ref{fig:fig6}).
Therefore, the applicability of the antibunching effect
for the Schmidt number estimation
is limited to
  weakly entangled states.

  In summary, our results elucidate the complex but structured relationship between spectral
  properties and quantum entanglement in biphoton states and their joint role in controlling
  two-photon interference signatures. Although theoretical analysis of this
    paper is restricted to the specific family of SPDC-biphotons,
  we hope it lays the groundwork for similar analytical treatments of broader families of biphoton
  states and

offers both conceptual
  insights and practical tools for the design of phase-sensitive quantum photonic systems. In
  particular, the demonstrated tunability of the antibunching response through spectral shaping
  opens new possibilities for phase-sensitive quantum metrology, with potential applications in
  compact, high-precision optical gyroscopes and related technologies.

\begin{acknowledgments}
The work of Guselnikov M. was financially supported by
the Gennady Komissarov Foundation for the Support of Young Researchers. Kiselev A. D. acknowledges support from the Russian Science
Foundation (Project No. 24-11-00398). Kozubov A. and Gaidash A. acknowledge support from the Russian Science
Foundation (Project No. 25-71-10034).
\end{acknowledgments}

\appendix

\section{Two-photon coincidence probability}\label{apendA0}

This appendix details the derivation of formula~\eqref{eq:eq2.1.11} which
  expresses the two-photon coincidence probability for biphotons in
  the state~\eqref{eq:eq2.1.3} brought to
  interfere at a balanced BS.

The probability of a two-photon coincidence event, $P_{2c}$, at the
output of the BS
can be evaluated by integrating the second-order
correlation function, $G^{(2)}$,
over the total time of data-taking
$t$ and detector time window $\tau$
as follows,
\begin{align}
P_{2c}=\int_{-\tau_f}^{\tau_f}d\tau\int_{-\infty}^{\infty}dt\ G^{(2)}(t,\tau),
\label{eq:eq2.1.5}
\end{align}
where $\tau_f$ is the width of the detector time window (the detector gate time) and the correlation function 
\begin{align}
  G^{(2)}(t,\tau) = &
  \bra{\Psi}\hat{U}_{BS}^{\dagger}\hat{a}^{\dagger}_1(t)\hat{a}^{\dagger}_2(t+\tau)
    \notag
  \\
  &
  \times
    \hat{a}_2(t+\tau)\hat{a}_1(t)\hat{U}_{BS}\ket{\Psi}, 
\label{eq:eq2.1.6}
\end{align}
is expressed in terms of
the Fourier transformed annihilation operators,
\begin{align}
\hat{a}_i(t) &= \frac{1}{\sqrt{2\pi}}\int_{-\infty}^{\infty}e^{-i\omega
             t}\hat{a}_i(\omega)d\omega,
\label{eq:eq2.1.8}
\end{align}
where $\hat{U}_{BS}$ is the operator of a balanced BS
  that acts on a creation operator as follows:
  $\hat{U}_{BS}a^\dagger_i(\omega)\hat{U}_{BS}^\dagger=\alpha_ia^\dagger_1(\omega)+\beta_ia^\dagger_2(\omega)$
  with the
  coefficients $\alpha_i$ and $\beta_i$ given by
  $\alpha_1=\beta_2=1/\sqrt{2}$ and $\beta_1=\alpha_2=i/\sqrt{2}$.

When describing the process of photodetection using
Fourier-transformed annihilation operators instead of field operators,
we treat radiation in terms of individual photons rather than as
an electromagnetic field. Within this framework, we apply narrow
bandwidth approximation, $\sigma_p\ll2\Omega$, and assume that the
photocurrent operator responsible for photon detection is proportional
to the Poynting vector, while the detector response time is much
shorter than the characteristic fluctuation timescale of the optical
signal. This formalism developed in~\cite{Blow1990} has been
applied to study HOM interference in Ref.~\cite{Toros2020, scott2020}.

We can now 
substitute the state~\eqref{eq:eq2.1.3} into Eq.~\eqref{eq:eq2.1.5}
and derive 
the coincidence probability
in 
the following form:
\begin{align}
  &
    P_{2c}=\int_{-\infty}^{\infty}\psi(\omega_1,
  \omega_2)\psi^*(\omega_1+\omega_2-\omega_2', \omega_2')\times
  \notag
  \\
&
  \left[\frac{\sin{[(\omega_2-\omega_2')\tau_f]}}{\omega_2-\omega_2'}-
  \frac{\sin{[(\omega_1-\omega_2')\tau_f]}}{\omega_1-\omega_2'}\right]\frac{d^3\omega}{2\pi}, 
\label{eq:eq2.1.9}
\end{align}
where $d^3\omega=d\omega_1d\omega_2d\omega_2'$.  We consider a state
  originating from SPDC so that its spectral amplitude $\psi(\omega_1,\omega_2)$ is proportional
  to the initial biphoton spectrum, which is non-negligible only within a narrow frequency region
  defined by $|\omega_1-\Omega|\sim\sigma_1$ and $|\omega_2-\Omega|\sim\sigma_2$. This means that
the integrand in  Eq.~\eqref{eq:eq2.1.9} is non-zero when $\omega_{1,2}-\omega_2'\sim\sigma_{1,2}$.
If the biphoton wave packet is significantly shorter than the detector gate time so that the
condition $\sigma_{1,2}\tau_f\gg1$ is fulfilled, the weak limit of the Dirac delta function
\begin{align}
\frac{\sin{[(\omega_{1,2}-\omega_2')\tau_f]}}{\omega_{1,2}-\omega_2'}\approx \pi\delta(\omega_{1,2}-\omega_2')
\label{eq:eq2.1.10}
\end{align}
can be applied to simplify Eq.~\eqref{eq:eq2.1.9}.
In practice,
typical values for the bandwidth of the biphoton state spectrum
$\sigma_{1,2}\sim10^{13}$~Hz and for the gate time 
$\tau_f\sim10^{-9}$~s~\cite{Cromb2023} can be used
to estimate the product $\sigma_{1,2}\tau_f$
at about $10^{4}$.
Numerical simulations show that
the latter is large enough to justify
the approximation~\eqref{eq:eq2.1.10}.

From Eqs.~\eqref{eq:eq2.1.10} and~\eqref{eq:eq2.1.9},
we obtain the well-known expression for
two-photon coincidence probability given by
\begin{align}
P_{2c}=\frac{1}{2}-\frac{1}{2}\int_{-\infty}^{\infty}d\omega_1d\omega_2\psi(\omega_1,
  \omega_2)\psi^*(\omega_2, \omega_1). 
\label{eq:eqAp0_2.1.11}
\end{align}
\section{Symmetry degree in the limit of
  vanishing pump variance}\label{append:sp=0}

In this appendix, we provide a detailed analysis of the symmetry degree
  \eqref{eq:eq2.2.2} in the limit $\sigma_p \to 0$. The key results of this analysis are summarized
  in Sec.~\ref{subsubsec:sigma-eq-0}.

  When the variance $\sigma_p$ approaches zero,
  $\sigma_p\to 0$,
  formulas~\eqref{eq:eq2.1.13} and~\eqref{eq:eq2.2.2}
can be used to simplify
the expression for the symmetry degree as follows
\begin{subequations}
  \label{eq:Ds_perft_ent}
\begin{align}
  &
    D_S=\frac{\int_{-\infty}^{\infty}\varphi_{12}(\tilde{\omega})\varphi^*_{12}(-\tilde{\omega})d\tilde{\omega}}{%
    \int_{-\infty}^{\infty}|\varphi_{1}(\tilde{\omega})|^2
    |\varphi_{2}(-\tilde{\omega})|^2d\tilde{\omega}}
    \label{eq:Ds_perft_phi}
  \\
  &
  =
  \frac{\int_{-\infty}^{\infty}\psi_{12}(\tilde{\omega})\psi^*_{12}(-\tilde{\omega})d\tilde{\omega}}{\int_{-\infty}^{\infty}
    |\psi_{12}(\tilde{\omega})|^2d\tilde{\omega}},
\label{eq:Ds_perft_psi}
\end{align}
\end{subequations}
where
\begin{align}
  \label{eq:psi12}
    \psi_{12}(\tilde{\omega})=\varphi_{1}(\tilde{\omega})\varphi_{2}(-\tilde{\omega}).
\end{align}
When $\varphi_{12}(\tilde{\omega})=\pm\varphi_{12}(-\tilde{\omega})$,
we have the symmetry degree~\eqref{eq:Ds_perft_phi}
in the form
\begin{align}
  &
    D_S\equiv D_{\pm}=\pm\frac{\int_{-\infty}^{\infty}|\varphi_{1}(\tilde{\omega})|^2
    |\varphi_{2}(\tilde{\omega})|^2d\tilde{\omega}}{%
    \int_{-\infty}^{\infty}|\varphi_{1}(\tilde{\omega})|^2
    |\varphi_{2}(-\tilde{\omega})|^2d\tilde{\omega}
    }.
\label{eq:Ds_perft_ent_plus_opc}
\end{align}
This formula
clearly demonstrates that the parity conditions~\eqref{eq:eq2.2.9}
do not guarantee the perfect (anti)bunching
with $D_{\pm}=\pm 1$.
The latter cannot be satisfied without imposing additional constraints.
From Eq.~\eqref{eq:Ds_perft_ent_plus_opc},
it is not difficult to see that the perfect (anti)bunching constraint
requires the magnitude of
either $\varphi_{1}(\tilde{\omega})$ or
$\varphi_{2}(\tilde{\omega})$ to be an even function of the frequency
\begin{align}
  &
    |\varphi_{i}(-\tilde{\omega})|^2=|\varphi_{i}(\tilde{\omega})|^2.
\label{eq:extra_cond_append}
\end{align}

  On the other hand,
  an alternative form of the perfect (anti)bunching condition
  can be obtained from
  relation~\eqref{eq:Ds_perft_psi}.
The result is given by
\begin{align}
  &
\psi_{12}(\tilde\omega)\equiv
    \varphi_1(\omega-\Omega)\varphi_2(\Omega-\omega)
    \notag
  \\
  &
    =
    \pm\varphi_1(\Omega-\omega)\varphi_2(\omega-\Omega)\equiv \pm\psi_{12}(-\tilde\omega).
\label{eq:perfect_antibunch_append}
\end{align}
It can be readily seen that, at $\tilde\omega_2=-\tilde\omega_1$, the JSA (anti)symmetry conditions,
$\psi(\tilde\omega_1,-\tilde\omega_1)=\pm\psi(-\tilde\omega_1,\tilde\omega_1)$,
applied to the spectral distribution~\eqref{eq:eq2.2.1}
are reduced to Eq.~\eqref{eq:perfect_antibunch_append}.
So, despite the fact that the latter has the form of a parity condition, it can be viewed as the limiting case of the (anti)symmetry conditions.

Importantly, though the parity conditions~\eqref{eq:eq2.2.9} combined with Eq.~\eqref{eq:extra_cond_append}
lead to Eq.~\eqref{eq:perfect_antibunch_append},
the perfect (anti)bunching and parity conditions are not equivalent.
In other words, Eq.~\eqref{eq:perfect_antibunch_append} generally
can be satisfied even if Eq.~\eqref{eq:eq2.2.9} does not hold.

  Note that, from Eq.~\eqref{eq:psi12},
  $\psi_{12}(\tilde\omega)$
  is invariant under the transformation:
  $\varphi_{1,2}(\tilde{\omega})\mapsto\psi_{1,2}(\tilde\omega)=\varphi_{1,2}(\tilde{\omega})\exp{\{f(\tilde{\omega})\}}$, where
  $f(-\tilde{\omega})=-f(\tilde{\omega})$ is an odd real-valued function,
  $\psi_{12}(\tilde\omega)=\varphi_{1}(\tilde{\omega})\varphi_{2}(-\tilde{\omega})=\psi_{1}(\tilde{\omega})\psi_{2}(-\tilde{\omega})$.
  When
  $\exp\{ 2 f(\tilde{\omega})\}=
|\varphi_{i}(-\tilde{\omega})/\varphi_{i}(\tilde{\omega})|
$,
  the squared amplitude of $\psi_i(\tilde\omega)$,
$|\psi_i(\tilde\omega)|^2=|\varphi_{i}(\tilde{\omega})\varphi_{i}(-\tilde{\omega})|$,
  meets the condition~\eqref{eq:extra_cond_append}
  which renders Eq.~\eqref{eq:perfect_antibunch_append}
  equivalent to the parity condition~\eqref{eq:eq2.2.9}
  with $\varphi_{12}=\varphi_1(\tilde\omega)\cnj{\varphi}_2(\tilde\omega)$ changed to $\psi_1(\tilde\omega)\cnj{\psi}_2(\tilde\omega)$.
 
\section{Scalar products}\label{apendA}
In this appendix,
we detail the derivation of formula~\eqref{eq:eq4.1.9}
for the scalar product that can be rewritten in the following form:
\begin{align}
  &
    \langle\phi_p|\tilde{\phi}_n \rangle=
    \notag
  \\
&\cos{(\beta\Omega)}\int_{-\infty}^{\infty}\psi_p\left(x\right)\psi_n\left(x\right)\cos{(\beta x
  s_1)}dx
  \notag
  \\
&-\sin{(\beta \Omega)}\int_{-\infty}^{\infty}\psi_p\left(x\right)\psi_n\left(x\right)\sin{(\beta x s_1)}dx,
\label{eq:eqA1}
\end{align}
where $x=(\omega_1-\Omega)/s_1$.
We can now use the known relation~\cite{Gradshteyn2014}
\begin{align}
  &\int_{-\infty}^{\infty}\psi_m\left(x\right)\psi_n\left(x\right)e^{i\sqrt{2}yx}dx
    \notag
  \\
  &
    =e^{-y^2/2}\frac{\sqrt{r!}}{\sqrt{s!}}(iy)^{s-r}L_r^{(s-r)}(y^2),
  \notag
  \\
&
  s=\max(m,n),\quad r = \min(m,n),
\label{eq:eqA2}
\end{align}
where $L_r^{(s-r)}(x)$ is the associated Laguerre polynomial,
to evaluate of the integrals that enter the right-hand side of Eq.~\eqref{eq:eqA1}.
The result reads
\begin{align}
  &
    \int_{-\infty}^{\infty}\psi_p\left(x\right)\psi_n\left(x\right)\cos{(\sqrt{2}xy)} dx
    \notag
  \\
  &
    =\frac{1+(-1)^{p+n}}{2} G_{pn}(y^2)e^{-y^2/2},
    \notag
  \\
  &
    \int_{-\infty}^{\infty}\psi_p\left(x\right)\psi_n\left(x\right)\sin{(\sqrt{2}xy)}dx
    \notag
  \\
  &
    =-i\frac{1-(-1)^{p+n}}{2}G_{pn}(y^2)e^{-y^2/2},
\label{eq:eqA3}
\end{align}
where $G_{pn}(x)=G_{np}(x)=\sqrt{r!/s!}(i\sqrt{x})^{s-r}L_r^{(s-r)}(x)$, $s=\max(p,n)$, $r=\min(p,n)$.
Note that $G_{nn}(x)=L_n(x)$ and $G_{pn}(0)=\delta_{pn}$.
For evaluation purposes,
$G_{mn}(x)$ can also be recast into the explicit polynomial form:
\begin{align}
  &
    G_{mn}(x)=(i\sqrt{x})^{|m-n|}\frac{(\frac{m+n+|m-n|}{2})!}{\sqrt{m!n!}}
    \notag
  \\
  &
    \times\sum_{k=0}^{(m+n-|m-n|)/2}
  % \left(\begin{matrix}\frac{m+n-|m-n|}{2}
%     \\
% k
%   \end{matrix}\right)
\binom{\frac{m+n-|m-n|}{2}}{k}
  \frac{(-1)^kx^k}{\left(|m-n|+k\right)!}.
\label{eq:eqA5}
\end{align}

Equation~\eqref{eq:eqA1} can now be combined with Eq.~\eqref{eq:eqA3}
to give formula~\eqref{eq:eq4.1.9}
\begin{align}
  \langle\phi_p|\tilde{\phi}_n \rangle =
  &
    \frac{e^{i\beta\Omega}+(-1)^{p+n}e^{-i\beta\Omega}}{2}
    \notag
  \\
  &
    \times G_{pn}\left(\frac{\beta^2 s_1^2}{2}\right)e^{-\beta^2 s_1^2/4}.
\label{eq:eqA6}
\end{align}
It is not difficult to check that
in accordance with Eq.~\eqref{eq:eqA1}
the expression~\eqref{eq:eqA6} is
real-valued and $\langle\phi_p|\tilde{\phi}_n \rangle=\langle\phi_p|\tilde{\phi}_n \rangle^*$.

Our concluding remark is that
calculations of the scalar product
$\langle\tilde{\phi}_n|\tilde{\phi}_m \rangle$
can be performed along similar lines.
The result is
\begin{align}
  &
    2 \langle\tilde{\phi}_n|\tilde{\phi}_m \rangle =\delta_{nm}
    \notag
  \\
  &
    +\cos{(2\beta\Omega)}\int_{-\infty}^{\infty}dx\psi_n\left(x\right)\psi_m\left(x\right)\cos{(2\beta
    x s_1)}
    \notag
  \\
  &
    -\sin{(2\beta
    \Omega)}\int_{-\infty}^{\infty}dx\psi_n\left(x\right)\psi_m\left(x\right)\sin{2\beta x
    s_1}=\delta_{nm}
    \notag
  \\
  &
    +\frac{e^{2i\beta\Omega}+(-1)^{n+m}e^{-2i\beta\Omega}}{2}G_{nm}(2\beta^2
    s_1^2)e^{-\beta^2 s_1^2}.
\label{eq:eqA7}
\end{align}

\section{Approximate Schmidt number for modulated biphoton state}
\label{apendB}

In this Appendix we discuss the derivation of
the approximate Schmidt number for
the cosine-modulated biphoton state~\eqref{eq:eq4.1.11}.
We begin with
the perturbation theory approach,
where the reduced density matrix~\eqref{eq:eq4.1.8} at $\beta = 0$
is considered as the unperturbed
density matrix $\hat{\rho}_1^{(0)}~=~\hat{\rho}_1\ (\beta~=~0)$
with the corresponding eigenvalues $\lambda_m^{(0)}=\lambda_m$.
We apply the first-order perturbation theory, assuming that small
deviations in the density matrix do not significantly alter the Schmidt modes.
Accordingly, $(\hat{\rho}_1^{(0)}+\varepsilon \hat{U})|\phi_m
\rangle\approx(\lambda_m^{(0)}+\varepsilon\lambda_m^{(1)})|\phi_m \rangle$,
where $\varepsilon$ is a small parameter and $\hat{U}$ is a first-order correction operator.
By assuming that $\hat{\rho}_1^{(0)}+\varepsilon \hat{U}\approx\hat{\rho}_1$ and
$\tilde{\lambda}_m\approx\lambda_m^{(0)}+\varepsilon\lambda_m^{(1)}$,
we can calculate the perturbed Schmidt value as follows:
\begin{align}
&\tilde{\lambda}_m\approx P_{cos}^{-1}\sum_{n=0}^{\infty} \lambda_n|\langle\phi_m|\tilde{\phi}_n \rangle|^2,
\label{eq:eqB1}
\end{align}
where the scalar product is given by Eq.~\eqref{eq:eqA6}.

In reality, the numerical evaluation of the approximate
Schmidt number using relation~\eqref{eq:eqB1}
is quite time-consuming and inconvenient.
We have found that the approximate $\tilde{\lambda}_m$ can be obtained through a
simpler and sufficiently accurate heuristic procedure.
Consider the action of the reduced density matrix
operator $\hat{\rho}_1$ on the state $|\tilde{\phi}_m \rangle_1$, which can be written in the form
\begin{align}
&\hat{\rho}_1|\tilde{\phi}_m \rangle_1 =P_{cos}^{-1}\lambda_m\langle\tilde{\phi}_m|\tilde{\phi}_m \rangle|\tilde{\phi}_m \rangle_1\notag\\
&+P_{cos}^{-1}\sum_{n\not=m} \lambda_n\langle\tilde{\phi}_n|\tilde{\phi}_m \rangle|\tilde{\phi}_n \rangle_1.
\label{eq:eqB2}
\end{align}
The scalar product $\langle\tilde{\phi}_n|\tilde{\phi}_m \rangle$ is given by Eq.~\eqref{eq:eqA7}.
So, we have
\begin{align}
  &
    \hat{\rho}_1|\tilde{\phi}_m \rangle_1 =\lambda_m\frac{ 1+e^{-\beta^2s_1^2}\cos{(2\beta\Omega)}
    L_m(2\beta^2s_1^2)}{1+e^{-\beta^2\alpha^2}\cos{(2\beta\Omega)}} |\tilde{\phi}_m \rangle
    \notag
  \\
  &
    +\frac{P_{cos}^{-1}}{2}e^{-\beta^2s_1^2}\sum_{n\neq
    m}\lambda_{n}\frac{e^{2i\beta\Omega}+(-1)^{n+m}e^{-2i\beta\Omega}}{2}
    \notag
  \\
  &
    \times G_{nm}(2\beta^2 s_1^2) |\tilde{\phi}_n \rangle.
\label{eq:eqB3}
\end{align}
In the unperturbed case limit, $\beta\rightarrow 0$, the sum in Eq.~\eqref{eq:eqB3} vanishes due to
$G_{nm}(0)=\delta_{nm}$. Moreover, the polynomial $G_{n\not=m}$ tends to zero faster in
$\beta^{|n-m|}$ times than the Laguerre polynomial tends to unity. So, in the first approximation,
we can omit the sum as follows:
\begin{align}
  &
    \rho_1|\tilde{\phi}_m \rangle_1 \approx\lambda_m\frac{ 1+e^{-\beta^2s_1^2}\cos{(2\beta\Omega)}
    L_m(2\beta^2s_1^2)}{1+e^{-\beta^2\alpha^2}\cos{(2\beta\Omega)}} |\tilde{\phi}_m \rangle
\label{eq:eqB4}
\end{align}
and obtain the following expression for the approximate Schmidt values:
\begin{align}
  &
    \tilde{\lambda}_m=\lambda_m\frac{ 1+e^{-\beta^2s_1^2}\cos{(2\beta\Omega)} L_m(2\beta^2s_1^2)}{1+e^{-\beta^2\alpha^2}\cos{(2\beta\Omega)}}.
\label{eq:eqB5}
\end{align}
Note that the Schmidt values normalization property is automatically satisfied for the
parameters~\eqref{eq:eqB2} and~\eqref{eq:eqB5} due to the form of the reduced density
matrix~\eqref{eq:eq4.1.8}.

The approximate heuristic Schmidt number is given by 
\begin{align}
&\tilde{K}=1/\sum_{n=0}^{\infty} \tilde{\lambda}_n^2.
\label{eq:eqB6}
\end{align}
The main challenge is to calculate the following sum:
\begin{align}
&\sum_{n=0}^{\infty} \lambda_n^2\Biggl(1+2e^{-\beta^2s_1^2}\cos{(2\beta\Omega)} L_n(2\beta^2s_1^2)\notag\\
&+e^{-2\beta^2s_1^2}\cos^2{(2\beta\Omega)} L_n^2(2\beta^2s_1^2)\Biggr).
\label{eq:eqB7}
\end{align}
The sum related to the first term in Eq.~\eqref{eq:eqB7} is merely the unperturbed ($\beta=0$) inverse Schmidt number \eqref{eq:eq3.1.4}:
\begin{align}
&\sum_{n=0}^{\infty} \lambda_n^2=K_0^{-1}.
\label{eq:eqB8}
\end{align}
The sum related to the second term in Eq.~\eqref{eq:eqB7} can be calculated with the help of the
generating function for the associated Laguerre polynomials~\cite{Bateman1953}:
\begin{equation}
\begin{aligned}
&\sum_{n=0}^{\infty}L^{(m)}_n(x)t^n = (1-t)^{-m-1}\exp{\left\{\frac{xt}{t-1}\right\}},
\label{eq:eqB9}
\end{aligned}
\end{equation}
leading to the result given by
\begin{align}
&\sum_{n=0}^{\infty} \lambda_n^2 L_n(2\beta^2s_1^2)=K_0^{-1}\exp{\left\{\frac{2\beta^2s_1^2z^4}{1-z^4}\right\}}.
\label{eq:eqB10}
\end{align}
The sum related to the third term in Eq.~\eqref{eq:eqB7} can be calculated using
the other generating function
for the products of the associated Laguerre polynomials~\cite{Bateman1953}:
\begin{align}
&\sum_{n=0}^{\infty}\frac{t^n n!}{\Gamma(n+m+1)}L^{(m)}_n(x)L^{(m)}_n(y) = \frac{1}{(xyt)^{m/2}(1-t)}\notag\\
&\times\exp{\left\{-(x+y)\frac{t}{1-t}\right\}}I_m\left(\frac{2\sqrt{xyt}}{1-t}\right),
\label{eq:eqB11}
\end{align}
where $\Gamma(x)$ is the Gamma function. 
The result
\begin{align}
  &\sum_{n=0}^{\infty} \lambda_n^2 L_n^2(2\beta^2s_1^2)=K_0^{-1}
    \notag
  \\
  &
    \times\exp{\left\{-4\beta^2s_1^2\frac{z^4}{1-z^4}\right\}}I_0\left(\frac{4\beta^2 s_1^2z^2}{1-z^4}\right)
\label{eq:eqB12}
\end{align}
can now be combined with
Eqs.~\eqref{eq:eqB8} and~\eqref{eq:eqB10} to give
the approximation~\eqref{eq:eq4.1.11} for the Schmidt number.

Remarkably, the heuristic Schmidt number given by Eq.~\eqref{eq:eq4.1.11} closely matches the
perturbative result obtained via Eq.~\eqref{eq:eqB1} in the regime of strong entanglement ($K>5$ or
$\sigma_p<0.1\sigma_1$).  In contrast, for weakly entangled states ($K<5$ or
$\sigma_p>0.1\sigma_1$), the heuristic expression proves to be significantly more accurate than the
first-order perturbation theory. Consequently, for modeling purposes, Eq.~\eqref{eq:eq4.1.11} offers
both practical and reliable alternatives.

\begin{figure*}
\centering
    \begin{subfigure}[b]{0.3\textwidth}
        \centering
        \includegraphics[width=\linewidth]{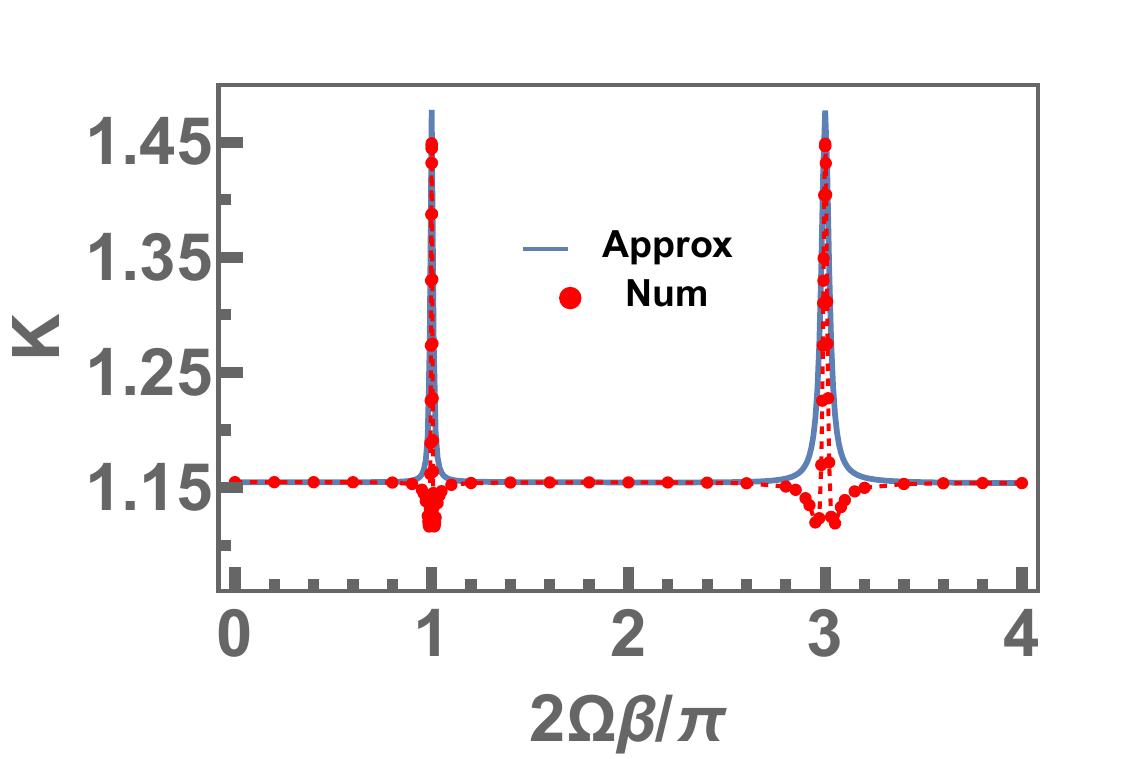}
        \caption{$\sigma_p=\sigma_1,~2\Omega\beta/\pi\in[0,4]$}
        \label{fig:fig7a}
    \end{subfigure}
   \hfill
    \begin{subfigure}[b]{0.3\textwidth}
        \centering
\includegraphics[width=\linewidth]{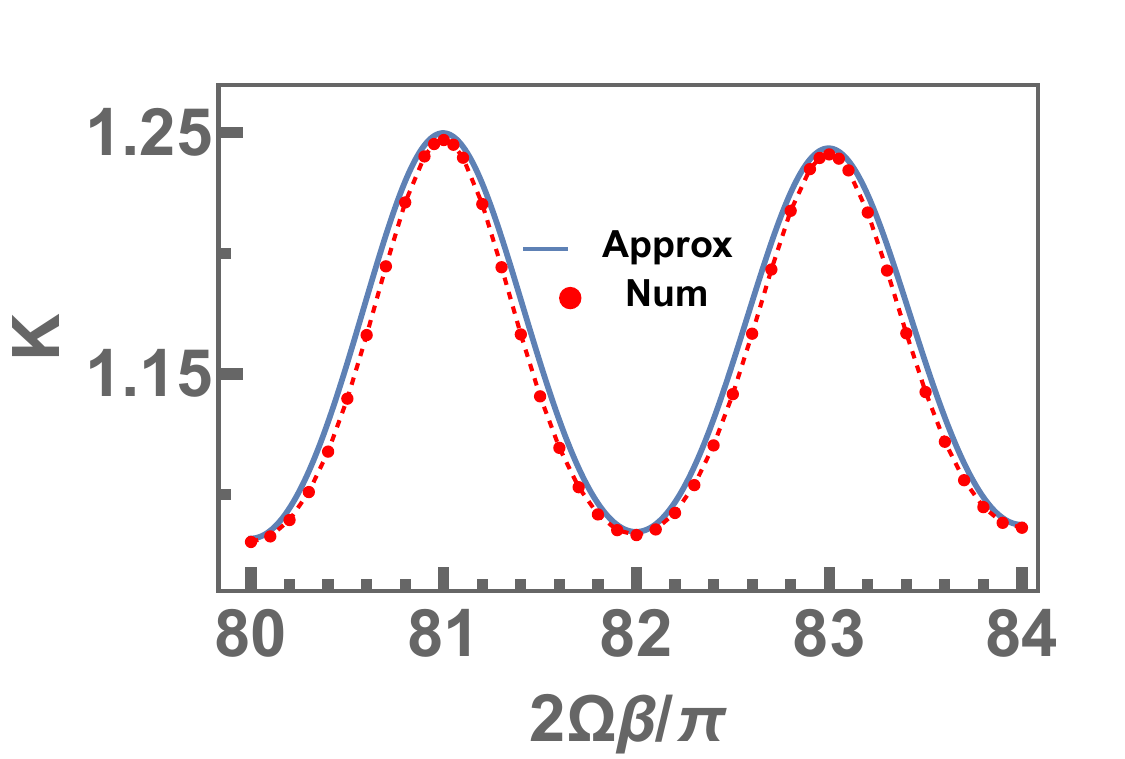}
        \caption{$\sigma_p=\sigma_1,~2\Omega\beta/\pi\in[80,84]$}
        \label{fig:fig7b}
    \end{subfigure}
    \hfill
    \begin{subfigure}[b]{0.3\textwidth}
        \centering
        \includegraphics[width=\linewidth]{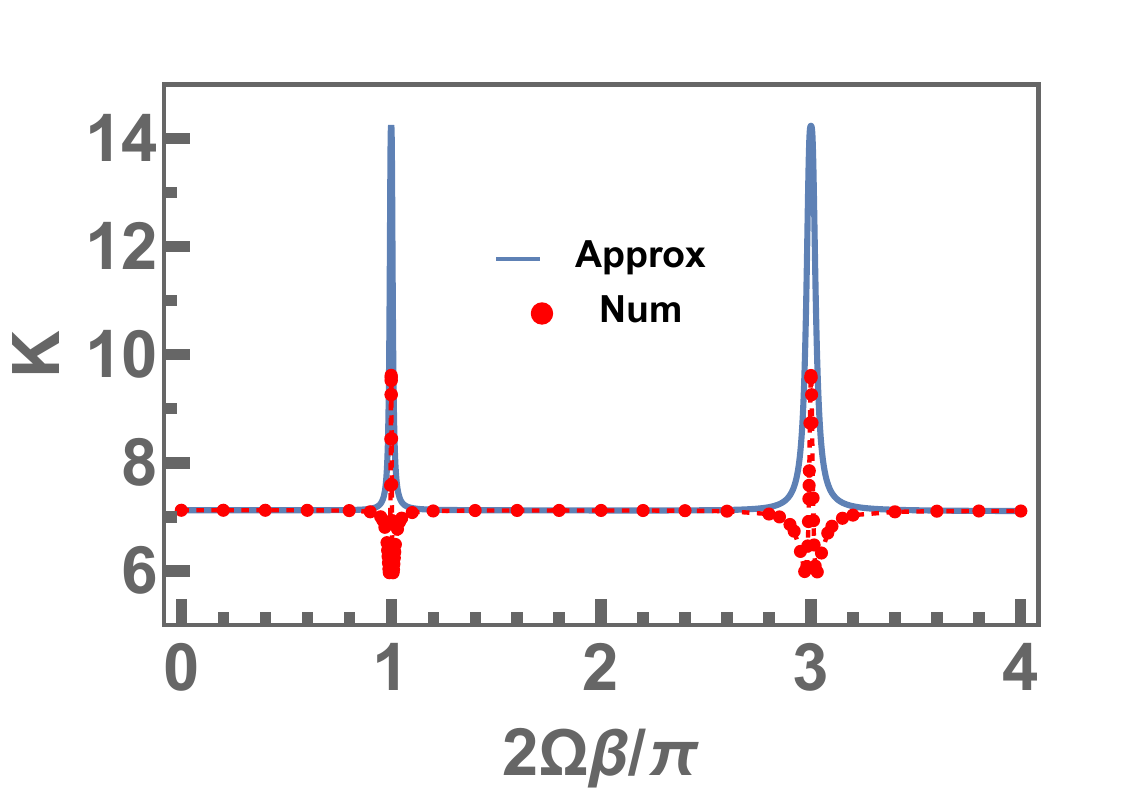}
        \caption{$\sigma_p=0.1\sigma_1,~2\Omega\beta/\pi\in[0,4]$}
        \label{fig:fig7c}
    \end{subfigure}
      \vspace{0.5em} 

    \begin{subfigure}[b]{0.3\textwidth}
        \centering
        \includegraphics[width=\linewidth]{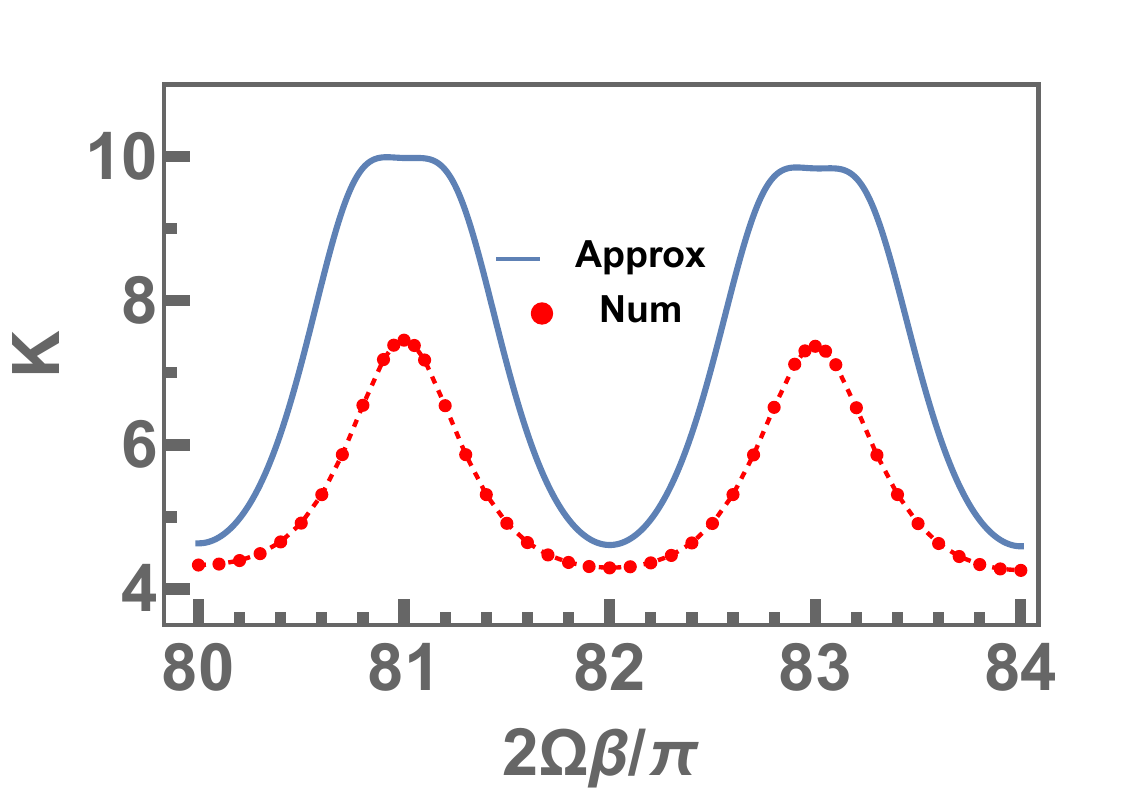}
        \caption{$\sigma_p=0.1\sigma_1,~2\Omega\beta/\pi\in[80,84]$}
        \label{fig:fig7d}
    \end{subfigure}
  \hfill
    \begin{subfigure}[b]{0.3\textwidth}
        \centering
        \includegraphics[width=\linewidth]{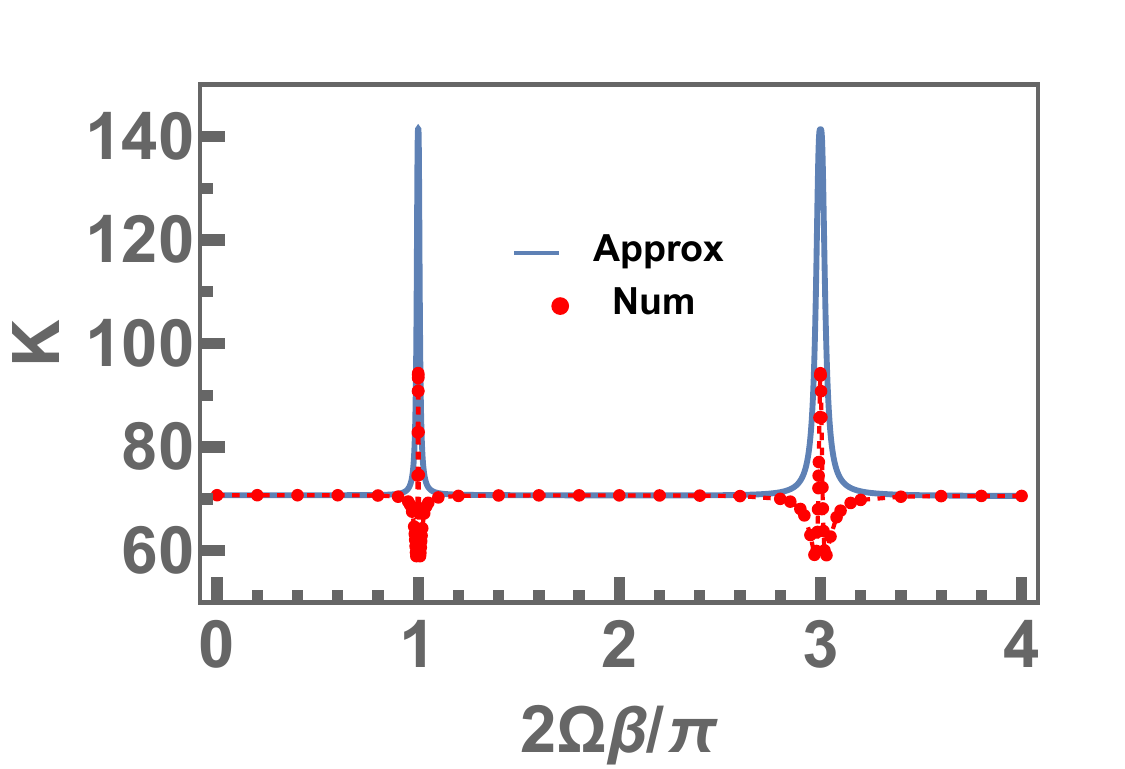}
        \caption{$\sigma_p=0.01\sigma_1,~2\Omega\beta/\pi\in[0,4]$}
        \label{fig:fig7e}
    \end{subfigure}
      \hfill
    \begin{subfigure}[b]{0.3\textwidth}
        \centering
        \includegraphics[width=\linewidth]{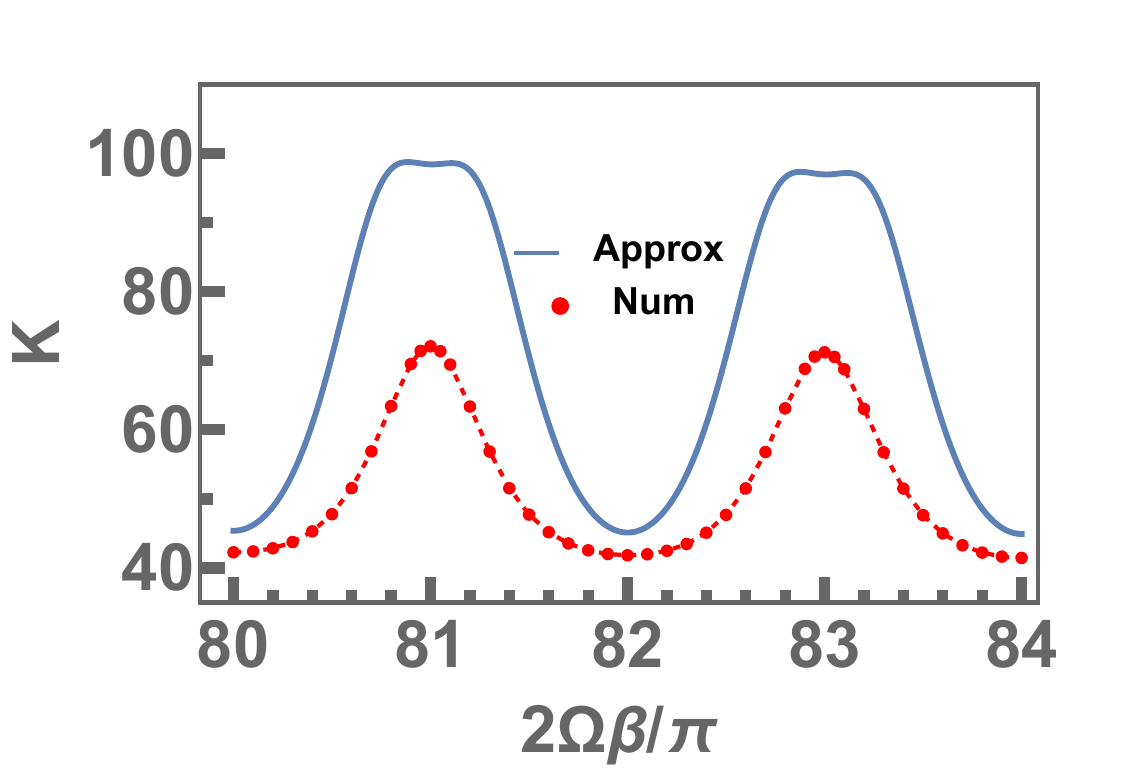}
        \caption{$\sigma_p=0.01\sigma_1,~2\Omega\beta/\pi\in[80,84]$}
        \label{fig:fig7f}
    \end{subfigure}

    \caption{ The Schmidt number, $K$, as a function of the time-delay parameter $\beta$ for the
      biphoton state with the cosine-modulated JSA.
Red dashed and blue solid curves represent the
      results of computing via the numerical calculations and the approximate formula~\eqref{eq:eq4.1.11},
      respectively.
      The
      parameters are listed in the caption of
      Fig.~\ref{fig:fig4}.}
\label{fig:fig7}
\end{figure*}

The results of comparing the approximate Schmidt number from Eq.~\eqref{eq:eq4.1.11} with numerical
calculations are shown in Fig.~\ref{fig:fig7}. Formula~\eqref{eq:eq4.1.11} provides a highly
accurate approximation of the Schmidt number for large $\sigma_p$ values; however, for small
$\sigma_p$, the approximation error increases. Nevertheless, the formula consistently predicts the
$\beta$ values corresponding to the minima and maxima of the entanglement degree with high accuracy,
providing a useful tool for the analytical analysis of the cosine-modulated biphoton state
entanglement degree.

%\nocite{*}

\bibliography{apssamp}% Produces the bibliography via BibTeX.

\end{document}